\documentclass[a4paper, 12pt]{article}
\pdfoutput=1
\usepackage{jheppub}
\usepackage[T1]{fontenc}
\usepackage{lmodern}
\usepackage{amsmath,amsfonts,amssymb,mathrsfs,mathtools,physics,tensor,graphicx,framed,float}
\usepackage{xcolor,tcolorbox}
\usepackage{caption}
\usepackage{subcaption}
\usepackage{tikz}
\usetikzlibrary{calc,arrows,decorations.markings}

\def\IZ {{\mathbb Z}}

\title{Topological vertex for 6d SCFTs with $\mathbb{Z}_2$-twist} 

\author[a,b]{Hee-Cheol Kim,}
\author[a]{Minsung Kim,}
\author[c]{Sung-Soo Kim}
\affiliation[a]{Department of Physics, POSTECH, \\
Pohang 37673, Korea}
\affiliation[b]{Asia Pacific Center for Theoretical Physics, \\ 
67 Cheongam-ro, Nam-gu, Pohang 37673, Korea}
\affiliation[c]{School of Physics, University of Electronic Science and Technology of China,\\
No. 2006 Xiyuan Ave, West Hi-Tech Zone, 
Chengdu, Sichuan 611731, China}

\abstract{
We compute the partition function for  6d $\mathcal{N}=1$ $SO(2N)$ gauge theories compactified on a circle with $\mathbb{Z}_2$ outer automorphism twist. We perform the computation based on 5-brane webs with two O5-planes using topological vertex with two O5-planes. As representative examples, we consider 6d $SO(8)$ and $SU(3)$ gauge theories with $\mathbb{Z}_2$ twist. We confirm that these partition functions obtained from the topological vertex with O5-planes indeed agree with the elliptic genus computations.
}
\begin{document}
\maketitle
\section{Introduction}\label{sec:introduction}
In \cite{Kim:2019dqn}, it was proposed that 5-brane webs \cite{Aharony:1997ju,Aharony:1997bh} for 6d $\mathcal{N}=(1,0)$ superconformal field theories (SCFTs) with $SO(N)$ gauge symmetry coupled to a tensor multiplet on a circle \cite{Heckman:2015bfa}. Such 5-brane webs are constructed with two O5-planes whose separation can be naturally identified with the Kaluza-Klein (KK) momentum \cite{Hayashi:2015vhy,Kim:2017jqn}. Given a 5-brane web, one can compute the prepotential \cite{Witten:1996qb,Intriligator:1997pq} for the corresponding theories on Coulomb branch. It was checked  \cite{Kim:2019dqn} that the prepotentials obtained from the proposed 5-branes webs indeed agree with the prepotentials that one can compute the triple intersection numbers in their geometric descriptions~\cite{Jefferson:2018irk,Bhardwaj:2019fzv}. 

Of particular significance to this 5-brane construction of 6d SCFTs with $SO(N)$ gauge symmetry is a realization of such 6d SCFTs with $\mathbb{Z}_2$ twist, providing a new perspective on RG flows on Higgs branches of D-type conformal matters \cite{Heckman:2013pva,DelZotto:2014hpa}. 
With non-zero holonomies turned on, one introduces a light charged scalar mode carrying non-zero KK momentum along the 6d circle,
giving rise to new Higgs branch associated with a vev of the light mode. 
In particular, RG flows from Higgsings on two O5-planes leads to  5-brane webs for twisted compactifications of 6d theories.

In this paper, we compute $\mathbb{R}^4\times T^2$ partition functions for 6d theories with $\mathbb{Z}_2$ twist, based on 5-brane configurations proposed in \cite{Kim:2019dqn}. As representative examples, we consider two 6d SCFTs with $\mathbb{Z}_2$ twist: one is the 6d $SO(8)$ gauge theory with $\mathbb{Z}_2$ twist and the other is the 6d $SU(3)$ theory with $\mathbb{Z}_2$ twist.  The $SO(8)$ theory is obtained through the Higgsing sequence from the 6d $SO(10)$ gauge theory with two hypermultiplets in the fundamental representation (flavors) to the $SO(9)$ gauge theory with a flavor and then to the $SO(8)$ theory with $\mathbb{Z}_2$ twist, while the $SU(3)$ theory is obtained through the Higgsing sequence from 6d $G_2$ gauge theory with a flavor to the $SU(3)$ theory with $\mathbb{Z}_2$ twist, whose 5-brane configuration yields 5d $SU(3)_9$ gauge theory with the Chern-Simons level $9$, as predicted in \cite{Jefferson:2017ahm,Jefferson:2018irk,Hayashi:2018lyv}.

 As a main tool of computation, we implement topological vertex formalism \cite{Aganagic:2003db,Iqbal:2007ii,Awata:2008ed} with an O5-plane \cite{Kim:2017jqn} applying these Higgsing sequences.  
 We check the obtained results against the elliptic genus~\cite{Haghighat:2013tka,Kim:2014dza, Haghighat:2014vxa,Gadde:2015tra}. computations by applying the Higgsings leading to the twisted theories or directly twisting.
 
 The organization of the paper is as follows. In section \ref{sec:SO8}, we discuss the construction of 5-brane webs for the 6d $SO(8)$ gauge theory with $\mathbb{Z}_2$ twist and compute the partition function using topological vertex based on 5-brane webs as well as using the ADHM method. In a similar way, the 6d $SU(3)$ gauge theory with $\mathbb{Z}_2$ twist is discussed in section \ref{sec:SU3}. We then summarize the results and discuss some subtle issues in section \ref{sec:conclusion}. In appendices \ref{sec:affine} and \ref{sec:special}, we discuss properties of $\mathbb{Z}_2$ twisting of 6d theories and the perturbative partition function from the perspective of twisted affine Lie algebras, and also provide various identities that are useful in actual computations.

\vspace{1cm}

While completing this paper, we became aware of \cite{Hayashi:2020hhb} which has some overlap with this paper.

\vspace{1cm}

\section{\texorpdfstring{$SO(8)$}{SO8} theory with \texorpdfstring{$\mathbb{Z}_2$}{Z2} twist} \label{sec:SO8}

We first consider twisted compactification of 6d pure $SO(8)$ gauge theory on $-4$ curve. As proposed in \cite{Kim:2019dqn}, a 5-brane configuration for the twisted compactification of the 6d pure $SO(8)$ gauge theory can be obtained from a Higgsing of 6d $SO(10)$ gauge theory with two hypermultiplets in the fundamental representation ($SO(10)+2\mathbf{F}$) on $-4$ curves. Before we discuss twisting, let us recall the standard Higgsing, which is to give the vev's to two fundamental scalars charged under a color D5-brane. This leads to the 6d $SO(8)$ gauge theory on $-4$ curve. 
To realize twisted compactification, one instead gives independent vev's to two individual scalars. Namely, first give a vev to one fundamental scalar 
to yield the 6d $SO(9)$ gauge theory with a fundamental hypermultiplet ($SO(9)+1\mathbf{F}$), and then give a different vev to another fundamental scalar carrying a unit KK momentum. This leads to the 6d $SO(8)$ gauge theory on $-4$ curve with $\IZ_2$ twist. In this section, we first review construction of 5-brane configuration leading to the twisted compactification of the 6d pure $SO(8)$ gauge theory on $-4$ curve, and then compute the partition function of the twisted $SO(8)$ gauge theory on $\mathbb{R}^4 \times T^2$, using topological vertex \cite{Aganagic:2003db} on a 5-brane web with two O5-planes \cite{Kim:2017jqn}. We compare our result with Higgsing or twisting of the elliptic genus from the ADHM construction using localization technique, introduced in \cite{Benini:2013nda, Benini:2013xpa}.

\begin{figure}[t]
\centering
\includegraphics[scale=1]{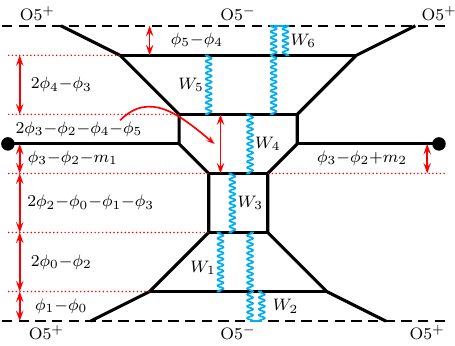}
\caption{A 5-brane web for 6d $SO(10)$ theory with two fundamentals. Two fundamental hypermultiplets are represented by two external D5-branes ending on a D7-brane (black dot) and the monodromy cuts of D7-branes point the outward directions.} \label{fig:6d_SO10}
\end{figure}
\subsection{5-brane web for 6d \texorpdfstring{$SO(8)$}{SO(8)} gauge theory with \texorpdfstring{$\mathbb{Z}_2$}{Z2} twist}
Let us begin with a 5-brane configuration for the 6d $SO(10)$ gauge theory with two fundamental hypermultiplets on $-4$ curve. It is depicted in Figure~\ref{fig:6d_SO10}, where there are two O5-planes whose separation naturally gives a compactification direction associated with 6d circle. Two fundamental hypermultiplets are denoted by two D5-branes ending on D7-branes. Their masses are $m_1, m_2$. The W-bosons $W_i$ of the theory are denoted by wiggly lines connecting color D5-branes in Figure~\ref{fig:6d_SO10}.
The masses of the W-bosons are given by
\begin{align}
 m_{W_1} &= 2\phi_0 - \phi_2\ ,& \quad m_{W_2} &= 2\phi_1 - \phi_2\ ,&  m_{W_3} &= 2\phi_2 - \phi_0 - \phi_1 - \phi_3\ , \nonumber \\
 m_{W_4} &= 2\phi_3 - \phi_2 - \phi_4 - \phi_5\ ,&  m_{W_5} &= 2\phi_4 - \phi_3\ ,& \quad 
 m_{W_6} &= 2\phi_5 - \phi_3\ ,
\end{align}
where $\phi_i$ ($i=1,\cdots,5$) are the Coulomb branch vev and $\phi_0$ is the tensor branch vev.  
As it is a configuration for the 6d theory on a circle, one can see that the W-bosons masses are consistent with the affine Cartan matrix\footnote{The affine Cartan matrix is defined as 
$\mathcal{C}_{ij} =\displaystyle 2\frac{(\alpha_i, \alpha_j)}{(\alpha_j, \alpha_j)}$ with the simple roots $\alpha_i$ of affine Lie algebras. 
} of untwisted affine Lie algebra $D_5^{(1)}$, 
\begin{align}
m_{W_{i}} = \big(\mathcal{C}_{ D_5^{(1)} }\big)_{ij}\, \phi_{j-1}\ ,\qquad \quad 
\mathcal{C}_{D_5^{(1)}}    = 
\begin{pmatrix*}[r]
 2 &  0 & -1 &  0 &  0 &  0\\ 
 0 &  2 & -1 &  0 &  0 &  0 \\
-1 & -1 &  2 & -1 &  0 &  0 \\
 0 &  0 & -1 &  2 & -1 & -1 \\
 0 &  0 &  0 & -1 &  2 &  0 \\
 0 &  0 &  0 & -1 &  0 &  2 
\end{pmatrix*}.
\end{align}

\begin{figure}
\centering
\begin{subfigure}[b]{0.45\textwidth}
\centering
\includegraphics[scale=1]{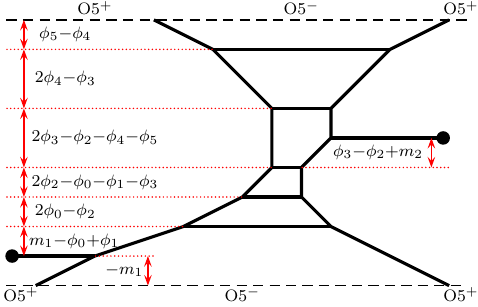}
\caption{}\label{fig:SO10_flop1a}
\end{subfigure}
\hfill
\begin{subfigure}[b]{0.5\textwidth}
\centering
\includegraphics[scale=1]{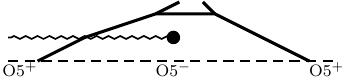}
\caption{}\label{fig:SO10_flop1b}
\includegraphics[scale=1]{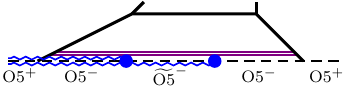}
\caption{}\label{fig:SO10_flop1c}
\includegraphics[scale=1]{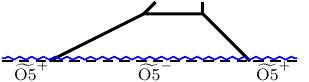}
\caption{}\label{fig:SO10_flop1d}
\end{subfigure}
\caption{(a) Another 5-brane web of 6d $SO(10)$ theory with two flavors after applying a series of flop transitions in Figure~\ref{fig:6d_SO10}. (b) The bottom part of (a) after applying Hanany-Witten transition. The black circle and black zigzag line denote D7-brane and corresponding monodromy cut, respectively. (c) Putting a D7-brane and a color D5-brane to the orientifold plane on the bottom. A D7-brane is split into two half D7-branes and generates $\widetilde{\mathrm{O}5}^-$ plane. The half D7-branes and corresponding half monodromy cuts are denoted by blue dots and blue zigzag lines. The half D5-brane is denoted by violet line. (d) A 5-brane configuration after Higgsing away two half D5-branes between D7-branes and then taking two half D7-branes to the left and right infinities, respectively.} \label{fig:SO10_flop1}
\end{figure}

\begin{figure}
\centering
\begin{subfigure}[b]{0.38\textwidth}
\centering
\includegraphics[scale=1]{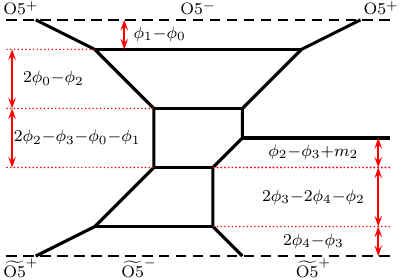}
\caption{}\label{fig:6d_SO9a}
\end{subfigure}
\hfill
\begin{subfigure}[b]{0.57\textwidth}
\centering
\includegraphics[scale=1]{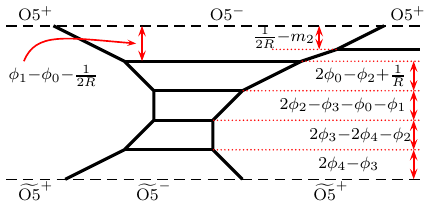}
\caption{}\label{fig:6d_SO9b}
\end{subfigure}
\caption{(a) The 5-brane web obtained by applying Figure~\ref{fig:SO10_flop1}(b)-(d) to the bottom part of Figure~\ref{fig:SO10_flop1}(a). (b) Flop transition of (a). } \label{fig:6d_SO9}
\end{figure}

As discussed, to twist, we perform the Higgsing such that we first bring a D7-brane down to one of the O5-planes to obtain $SO(9)$ gauge theory with a fundamental as depicted in Figure~\ref{fig:SO10_flop1}, and then push up the remaining D7-brane to the other O5-plane to give a different vev to the remaining scalar, as in Figure~\ref{fig:6d_SO9}.
More precisely, the Higgsing from a $SO(10)$ gauge theory with a flavor to a $SO(9)$ gauge theory is achieved as follows: Through successive flop transitions, one can bring a D7-brane near the bottom O5-plane as depicted in Figure~\ref{fig:SO10_flop1}(\subref{fig:SO10_flop1a}). Then one brings the D7-brane inside the Coulomb branch of $SO(10)$ gauge theory, where the D7-brane becomes floating as in Figure~\ref{fig:SO10_flop1}(\subref{fig:SO10_flop1b}), since the flavor D5-brane is annihilated due to the Hanany-Witten transition. We give a vev to the flavor which locates the flavor 7-brane and a color D5-brane on an O5-plane. That is to set the masses of the flavor hypermultiplet and the associated W-boson $\phi_1-\phi_0$ to zero,   
\begin{align}\label{eq:SO8Z2_higgsing1}
m_1 = 0\ , \qquad \phi_1 = \phi_0\ .
\end{align}

We note that when it is placed on an O5-plane as depicted in Figure \ref{fig:SO10_flop1}(\subref{fig:SO10_flop1b}) and Figure \ref{fig:SO10_flop1}(\subref{fig:SO10_flop1c}), a D7-brane (black dot) is split into two half D7-branes (blue dots), creating a half D5-brane between two half D7-branes stuck on an O5-plane, which hence turns the O5-plane  between these split half D7-branes to an $\widetilde{{\rm O5}}^-$-plane \cite{Evans:1997hk, Giveon:1998sr, Feng:2000eq, Bertoldi:2002nn}.
In doing so, a new Higgs branch opens up such that two half color D5-branes (or a full color D5-brane) are suspended between two half D7-branes on the $\widetilde{{\rm O5}}^-$-plane which can be Higgsed away, resulting in a 5-brane configuration for a $SO(9)$ gauge theory. 
After moving half D7-branes away from each other to the opposite 
along the orientifold plane, one gets a 5-brane configuration with an $\widetilde{{\rm O5}}$-plane as in Figure~\ref{fig:SO10_flop1}(\subref{fig:SO10_flop1d}).  

Figure~\ref{fig:6d_SO9}(\subref{fig:6d_SO9a}) is the resulting 5-brane web diagram describing the 6d $SO(9)$ gauge theory on a circle with a fundamental hypermultiplet of mass $m_2$, implementing \eqref{eq:SO8Z2_higgsing1}. 
Here we have redefined the scalars $\phi_i$ 
\begin{align}
      \phi_0 \to \phi_4\ ,
\qquad \phi_2 \to \phi_3\ , 
\qquad \phi_3 \to \phi_2\ , 
\qquad \phi_4 \to \phi_0\ , 
\qquad \phi_5 \to \phi_1\ ,
\end{align}
to make the masses of the W-bosons form the affine Cartan matrix of untwisted affine Lie algebra $B_4^{(1)}$, 
\begin{align}
m_{W_1} &= 2\phi_0 - \phi_2\ ,&  
m_{W_2} &= 2\phi_1 - \phi_2\ ,&    
m_{W_3} &= 2\phi_2 -\phi_3- \phi_0 - \phi_1\ , \nonumber \\
m_{W_4} &= 2\phi_3 - 2\phi_4-\phi_2\ , &
m_{W_5} &= 2\phi_4 - \phi_3\ ,& 
\end{align}
or more explicitly,
\begin{align}
m_{W_{i}} = \big(\mathcal{C}_{B_4^{(1)}}\big)_{ij}\, \phi_{j-1}\ ,\qquad \quad 
\mathcal{C}_{B_4^{(1)}}    = 
\begin{pmatrix*}[r]
 2 &  0 & -1 &  0 &  0 \\ 
 0 &  2 & -1 &  0 &  0 \\
-1 & -1 &  2 & -1 &  0 \\
 0 & 0 &  -1 &  2 & -2 \\
 0 &  0 &  0 & -1 &  2 
\end{pmatrix*}.
\end{align}

We further Higgs the remaining fundamental hypermultiplet. As the 5-brane configuration given in Figure \ref{fig:6d_SO9} has two different types of orientifold planes, O5- and  $\widetilde{\rm O5}$-planes,  we have two ways of Higgsing the remaining hypermultiplet. In the previous Higgsing $SO(10)+2\mathbf{F}\to SO(9)+1\mathbf{F}$, the Higgsing procedure converts an O5-plane to an $\widetilde{\rm O5}$-plane or vice versa. In a similar fashion, if one brings a flavor D5-brane (or equivalently a flavor D7-brane) down to the $\widetilde{\rm O5}$-plane, then it is to perform the same Higgsing, and hence it would yield a 5-brane web for a pure $SO(8)$ theory on a circle. This is the standard Higgsing $ SO(9)+1\mathbf{F}\to SO(8) $, hence corresponding to untwisted compactification $SO(10)+2\mathbf{F}\to SO(9)+1\mathbf{F} \to SO(8)$.
Indeed, shifting $\phi_3
\to \phi_3+\phi_4$ straightforwardly yields the W-boson masses forming the affine Cartan matrix of untwisted affine Lie algebra $D_4^{(1)}$, as expected, 
\begin{align}
m_{W_1} &= 2\phi_0 - \phi_2\ ,&  
m_{W_2} &= 2\phi_1 - \phi_2\ ,&    
m_{W_3} &= 2\phi_2 -\phi_3-\phi_4- \phi_0 - \phi_1\ , \nonumber \\
m_{W_4} &= 2\phi_3 -\phi_2\ , &
m_{W_5} &= 2\phi_4 - \phi_2\ .& 
\end{align}

\begin{figure}
\centering
\includegraphics[scale=1]{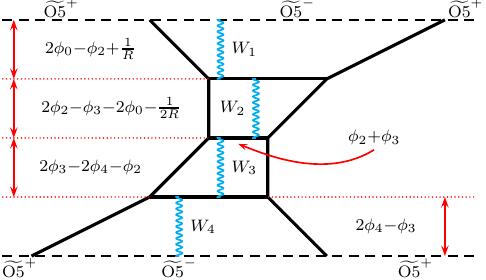}
\caption{The 5-brane web of $SO(8)$ theory with $\mathbb{Z}_2$ twist.} \label{fig:SO8Z2}
\end{figure}

Now, let us take the flavor D5-brane and one of color D5-branes to the ${\rm O5}$-plane instead. One can do the same kind of Higgsing as done for the previous the $SO(10)$ theory to the $SO(9)$ theory, which then makes a 5-brane configuration with two $\widetilde{\rm O5}$-planes. As discussed in \cite{Kim:2019dqn}, 
this is the Higgsing that one Higgses
the $SO(9)$ theory with one fundamental hypermultiplet by giving a vev to a scalar field carrying Kaluza-Klein momentum which leads to the $SO(8)$ gauge theory with a $\mathbb{Z}_2$ twist. Namely, we first revive the 6d circle radius $R$ in Figure \ref{fig:6d_SO9}(\subref{fig:6d_SO9a}) by introducing Kaluza-Klein momentum, which can be done by redefining K\"ahler parameters to be 
\begin{align}
2\phi_0 - \phi_2 \to 2\phi_0 - \phi_2 + \frac{1}{R}\ , \quad\qquad  \phi_1 - \phi_0 \to \phi_1 - \phi_0 - \frac{1}{2R}\ ,
\end{align}
where other parameters are unaltered in Figure~\ref{fig:6d_SO9}(\subref{fig:6d_SO9a}). By successive flop transitions, the fundamental hypermultiplet can be placed near the O5-plane located on the top in Figure~\ref{fig:6d_SO9}(\subref{fig:6d_SO9a}). This yields a 5-brane configuration in Figure~\ref{fig:6d_SO9}(\subref{fig:6d_SO9b}). We then Higgs by setting
\begin{align}\label{eq:SO8Z2_higgsing2}
m_2 = \frac{1}{2R}\ , \qquad \qquad \phi_1 = \phi_0 + \frac{1}{2R}\ .
\end{align}
The resulting 5-brane configuration for the 6d $SO(8)$ gauge theory with a $\mathbb{Z}_2$ twist is depicted in Figure \ref{fig:SO8Z2}. Here the W-boson masses are given by
\begin{align}
m_{W_1} &= 2\phi_0 - \phi_2 + \frac{1}{R}\ , 
&m_{W_2} &= 2\phi_2 - 2\phi_0 - \phi_3 - \frac{1}{2R} \ ,\nonumber \\
 m_{W_3} &= 2\phi_3 - 2\phi_4 - \phi_2\ , 
 &m_{W_4} &= 2\phi_4 - \phi_3\ ,
\end{align}
which form the affine Cartan matrix of twisted affine Lie algebra $D_4^{(2)}$,
\begin{align}
\mathcal{C}_{D_4^{(2)}} =\left(\begin{array}{rrrr}
2 & -1 & 0 & 0 \\
-2 & 2 & -1 & 0 \\
0 & -1 & 2 & -2 \\
0 & 0 & -1 & 2
\end{array}\right).
\end{align}


\subsection{Partition function from 5-brane webs}
We now compute the partition function based on the 5-brane configuration for the 6d $SO(8)$ gauge theory with a $\mathbb{Z}_2$ twist, using topological vertex with an O5-plane \cite{Kim:2017jqn}. 
With an O5-plane, we can easily realize a 5d $SO(N)$ or $Sp(N)$ gauge theory with hypermultiplets in the fundamental representation. For $SO(N)$ gauge theories, we can also depict 5-brane configurations for hypermultiplets in the spinor representation~\cite{Zafrir:2015ftn}. Our strategy to carry out topological vertex computation with O5-planes is first to introduce auxiliary spinor hypermultiplets and then to decouple them after the computation.

\begin{figure}
\centering
\begin{subfigure}{0.99\textwidth}
\centering
\includegraphics[scale=1]{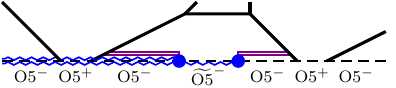}
\caption{}\label{fig:spinora}
\end{subfigure}
\begin{subfigure}{0.45\textwidth}
\centering
\includegraphics[scale=1]{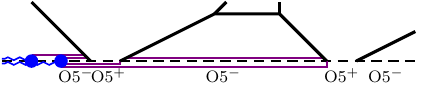}
\caption{}\label{fig:spinorb}
\end{subfigure}
\hfill
\begin{subfigure}{0.45\textwidth}
\centering
\includegraphics[scale=1]{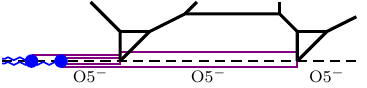}
\caption{}\label{fig:spinorc}
\end{subfigure}
\caption{(a) Couple two spinors after Higgsing Figure~\ref{fig:SO10_flop1}(c). (b) Moving two half D7-branes and corresponding monodromy cuts to far left. (c) Applying generalized flop transition to spinors.} \label{fig:spinor}
\end{figure}
\begin{figure}
\centering
\begin{subfigure}{0.45\textwidth}
\centering
\includegraphics[scale=1]{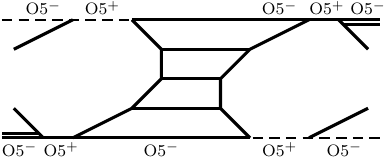}
\caption{}\label{fig:6d_SO8Z2_spinor-a}
\end{subfigure}
\begin{subfigure}{0.45\textwidth}
\centering
\includegraphics[scale=1]{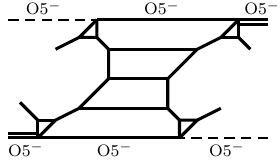}
\caption{}\label{fig:6d_SO8Z2_spinor-b}
\end{subfigure}
\caption{A 5-brane web diagram for $\mathbb{Z}_2$ twisted compactification of 6d $SO(8)$ theory with four auxiliary spinors.} \label{fig:SO8+4S}
\end{figure}

To introduce four auxiliary spinors, we start with a 5-brane configuration before twisting. Namely, we consider 5-brane web for 6d $SO(10)$ gauge theory with two hypermultiplets in the fundamental representation as well as four auxiliary hypermultiplets in the spinor representation. One can imagine that given a 5-brane web for 6d $SO(10)$ gauge theory with two fundamentals as in Figure~\ref{fig:SO10_flop1}(\subref{fig:SO10_flop1a}), one introduces each spinor as a charge conserving distant 5-brane on the left and on the right of the main web for the $SO(10)$ gauge theory. On the bottom O5-plane, they are a 5-brane of charge $(2, -1)$ on the left and a 5-brane of charge $(2, 1)$ on the right. On the top O5-plane, they are a 5-brane of charge $(2, -1)$ on the left and a 5-brane of charge $(2,-1)$ on the right. Higgsing toward $SO(8)$ gauge theory with a twist is straightforward except there are additional half D5-branes introduced along O5-planes due to Hanany-Witten transitions when taking half D7-brane to infinity. For instance, in Figure~\ref{fig:spinor}, Higgsing from $SO(10)$ gauge theory with a fundamental is depicted in the presence of two spinors, a distant 5-brane of charge $(1,-1)$ due to the monodromy of D7-brane (fundamental hyper) on the left and another distant 5-brane of charge $(2,1)$ on the right of the bottom O5-plane.  Figure~\ref{fig:spinor}(\subref{fig:spinora}) is a configuration when the Higgsing is performed, where there are two half D7-branes are on the orientifold plane. Unlike the previous case where we took each half D7-brane to the opposite directions to realize an $\widetilde{\rm O5}$-plane, this time, for computation ease, we take two half D7-branes to the same direction, to the left, to keep the orientifold plane an O5-plane. This is depicted in Figure~\ref{fig:spinor}(\subref{fig:spinorb}), where half D5-branes are created due to the Hanany-Witten transitions. Taking the spinor 5-branes closer to the center of the O5-plane, they undergo ``generalized flop'' transitions~\cite{Hayashi:2017btw} but still preserve the charge conservation as giving in Figure~\ref{fig:spinor}(\subref{fig:spinorc}). Repeating the same procedure on the top O5-plane, one finds a 5-brane web for the 6d $SO(8)$ gauge theory with four auxiliary spinors with a $\mathbb{Z}_2$ twist, as in Figure~\ref{fig:SO8+4S}. 
Again, to obtain the partition function for the 6d $SO(8)$ gauge theory with a $\mathbb{Z}_2$ twist, we first compute the partition function based on the 5-brane web for the 6d $SO(8)$ gauge theory with four spinors with the twist in  Figure~\ref{fig:SO8+4S}, and then decouple the spinors by taking all the masses of the spinor matter to infinity.

\begin{figure}
\centering
\includegraphics[scale=1]{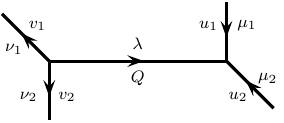}
\caption{The example of 5-brane web where $\lambda, \mu_i, \nu_i$ are Young diagram associated with edges, $Q$ is a K\"ahler parameter associated with edge $\lambda$, and $u_i, v_i$ are $(p, q)$-charges of four edges.} \label{fig:vtx_method}
\end{figure}

To perform the topological vertex method with an O5-plane, we assign arrows, Young diagrams $\lambda, \mu_i, \nu_i$ and K\"ahler parameters $Q_i$ on the edges as in Figure~\ref{fig:vtx_method}. The topological string partition function $Z$ can be computed by evaluating the edge factor and the vertex factor: 
\begin{align}
Z = \sum_{\lambda,\mu,\nu,\cdots} \qty(\prod \mathrm{Edge\:Factor}) \qty( \prod \mathrm{Vertex\:Factor})\ .
\end{align}
The Edge Factor is given by the product of K\"ahler parameters and the framing factor of the associated Young diagram $\lambda=(\lambda_{1},\lambda_{2},\cdots, \lambda_{\ell(\lambda)})$, 
\begin{align}
    (-Q)^{|\lambda|}f^n_\lambda\ ,
\end{align}
where the framing factor is defined as 
\begin{align}\label{eq:framing_def}
f_\lambda = (-1)^{\abs{\lambda}} g^{\frac{1}{2}(\norm*{\lambda^t}^2 - \norm*{\lambda}^2)} = f_{\lambda^t}^{-1}\ ,
\end{align}
with  
\begin{align}
\abs{\lambda} = \sum_{i=1}^{\ell(\lambda)} \lambda_{i}\ , \qquad
\norm{\lambda}^2 = \sum_{i=1}^{\ell(\lambda)} \lambda_{i}^2\ .
\end{align}
Here, 
$g= e^{- \epsilon}$ 
is the $\Omega$-deformation parameter that is unrefined $\epsilon_1=-\epsilon_2=\epsilon$, and $\lambda^t$ denotes the transposed Young diagram which corresponds to the conjugate partition of $\lambda$. The power of the framing factor $n$ is defined with the incoming and outgoing edges connected to $Q$ and their antisymmetric product, for instance, in Figure~\ref{fig:vtx_method}, $n= u_i \wedge v_2= \det(u_1, v_2)$.

The Vertex Factor is given by 
\begin{align}\label{eq:vtx_def}
C_{\lambda \mu \nu} = g^{\frac{1}{2}(-\norm*{\mu^t}^2 + \norm*{\mu}^2 + \norm*{\nu}^2)} \tilde{Z}_{\nu}(g) \sum_{\eta} s_{\lambda^t/\eta}(g^{-\rho-\nu}) s_{\mu/\eta}(g^{-\rho-\nu^t})\ ,
\end{align}
where $\lambda, \mu, \nu $ are the Young diagrams assigned to three out-going edges connected to a given vertex, arranged in the clockwise order such that the last Young diagram $\nu$ is assigned to the preferred direction which is associated with instanton sum. We note that $C_{\lambda \mu \nu}=C_{\mu \nu\lambda }=C_{\nu\lambda \mu }$ for unrefined vertex factors. When the orientation of an edge is reversed, the corresponding Young diagram is transposed.
Here, $\tilde{Z}_{\nu}(g)$ is defined by
\begin{align}
    \tilde{Z}_\nu(g) &= \tilde{Z}_{\nu^t}(g)
= \prod_{i=1}^{\ell(\nu)} \prod_{j=1}^{\nu_i} \qty(1 - g^{\nu_i + \nu_j^t - i - j + 1})^{-1},
\end{align}
where $s_{\lambda/\eta}(x)$ is skew Shur functions and $\rho=(-\frac{1}{2}, -\frac{3}{2}, -\frac{5}{2}, \cdots)$.  When summing skew Shur functions, one needs to repeatedly use the Cauchy identities 
\begin{align}
&\sum_\lambda Q^{\abs{\lambda}} s_{\lambda/\eta_1}(g^{-\rho-\nu_1}) s_{\lambda/\eta_2}(g^{-\rho-\nu_2}) \nonumber \\
&\quad = \mathcal{R}_{\nu_2 \nu_1}(Q)^{-1} \sum_\lambda Q^{\abs{\eta_1}+\abs{\eta_2}-\abs{\lambda}} s_{\eta_2/\lambda}(g^{-\rho-\nu_1}) s_{\eta_1/\lambda}(g^{-\rho-\nu_2}) \label{eq:cauchy2}\ , \\
&\sum_\lambda Q^{\abs{\lambda}} s_{\lambda/\eta_1^t}(g^{-\rho-\nu_1}) s_{\lambda^t/\eta_2}(g^{-\rho-\nu_2}) \nonumber \\
&\quad = \mathcal{R}_{\nu_2 \nu_1}(-Q) \sum_\lambda Q^{\abs{\eta_1}+\abs{\eta_2}-\abs{\lambda}} s_{\eta_2^t/\lambda}(g^{-\rho-\nu_1}) s_{\eta_1/\lambda^t}(g^{-\rho-\nu_2}) \ \label{eq:cauchy2'},
\end{align}
where  
\begin{align}
     \mathcal{R}_{\lambda \mu}(Q)
&= \prod^{\infty}_{i,j=1}\left(1- Q\, g^{i+j-\mu_i-\lambda_j-1}\right).
\end{align}
See also Appendix~\ref{sec:special} for various other Cauchy identities and other related special functions.

\begin{figure}
\centering
\includegraphics[scale=1]{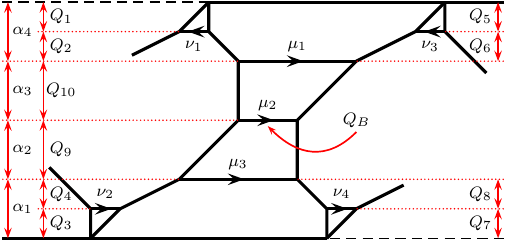}
\caption{5-brane web diagram for $SO(8)$ theory with a $\mathbb{Z}_2$ twist where we first couple four auxiliary spinors and then later decouple all the spinors after obtaining the partition function. In this way, one can compute the partition function for 6d $SO(8)$ theory with a $\mathbb{Z}_2$ twist.} 
\label{fig:SO8Z2_vtx}
\end{figure}

\begin{figure}
\centering
\begin{subfigure}{0.45\textwidth}
\centering
\includegraphics[scale=1]{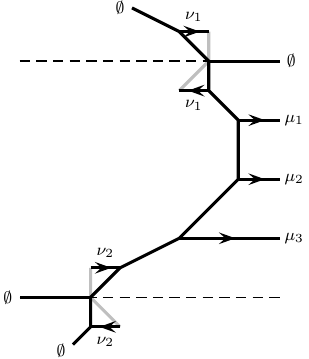}
\caption{}
\label{fig:SO8Z2_vtx_La}
\end{subfigure}
\begin{subfigure}{0.45\textwidth}
\centering
\includegraphics[scale=1]{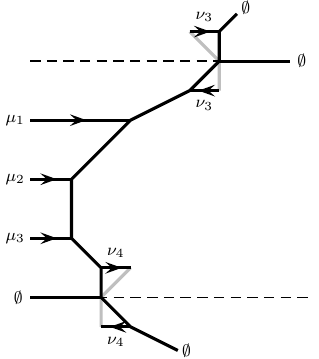}
\caption{}
\label{fig:SO8Z2_vtx_Lb}
\end{subfigure}
\caption{Strip diagrams for 6d $SO(8)$ gauge theory with $\mathbb{Z}_2$ twist, which is coupled to 4 auxiliary spinors. They are the left part (a) and the right part (b) of Figure~\ref{fig:SO8Z2_vtx}. We use mirror images to make each part as a form of a strip diagram.} 
\label{fig:SO8Z2_vtx_LR}
\end{figure}

Now, we assign the Young diagrams and K\"ahler parameters to a 5-brane web diagram for the 6d $SO(8)$ gauge theory with four auxiliary spinors as in Figure~\ref{fig:SO8Z2_vtx}. We compute the partition function based on this 5-brane web with two O5-planes as follows: First we reflect the 5-brane web with respect to each O5-plane, which creates a mirror image of the original 5-brane web. We also cut the 5-brane configuration by half so that we can glue over the color D5-branes associated with Young diagrams $\mu_1, \mu_2,\mu_3$. As we have mirror images, we can choose a convenient fundamental 5-brane configuration, which contains both the original and the mirror images such that when crossing the O5-plane, 5-brane webs are smoothly connected.  
Together with the mirror images, we then see that we can extract two pieces of strip diagrams given in Figure~\ref{fig:SO8Z2_vtx_LR}.
We note here in Figure~\ref{fig:SO8Z2_vtx_LR}  that 
unlike the edge factors $\mu_i$ following the conventional rule, the edge factors of $\nu_i$ should be treated carefully. When $\nu_i$ are reflected over the orientifold planes, they get the additional factor $(-1)^{\abs{\nu_i}} f_{\nu_i}^{\pm 1}$, where $\pm 1$ is determined from the orientation of the edges. For the vertex factor, the assigned arrows associated with Young diagrams are reversed when they are reflected over, which makes the ordering of the corresponding Young diagrams counterclockwise. It also follows from the identity~\cite{Kim:2017jqn}
\begin{align}\label{eq:Ct}
C_{\lambda \mu  \nu} 
= (-1)^{| \lambda | + | \mu | + | \nu |} 
f_{\lambda}^{-1}(g) f_{\mu}^{-1}  (g) f_{\nu}^{-1}  (g)
C_{\mu^t \lambda^t \nu^t}\ ,
\end{align}
that the reversal of the Young diagrams is translated into the transposition of the Young diagrams, and in turn, it is equivalent to changing the direction of the arrows. 

The topological string partition function is then  obtained by gluing two strip diagrams in Figure \ref{fig:SO8Z2_vtx_LR}, by summing up the Young diagrams $\mu_i$ and $\nu_i$ which are associated with the preferred direction
\begin{align}\label{eq:SO8Z2_vtx}
Z &= \sum_{\mu_i, \nu_i} Q_1^{\abs{\nu_1}} Q_3^{\abs{\nu_2}} Q_5^{\abs{\nu_3}} Q_7^{\abs{\nu_4}} (-Q_B Q_{10})^{\abs{\mu_1}} (-Q_B)^{\abs{\mu_2}} (-Q_B Q_{9})^{\abs{\mu_3}} \nonumber \\
& \qquad \quad \times f_{\mu_1}^2 f_{\mu_3}^{-2} f_{\nu_1}^3 f_{\nu_2} f_{\nu_3} f_{\nu_4}^3 Z^{\rm strip}_{\mathrm{left}}(\mu_1, \mu_2, \mu_3, \nu_1, \nu_2) Z^{\rm strip}_{\mathrm{right}}(\mu_1, \mu_2, \mu_3, \nu_3, \nu_4)\ , 
\end{align}
where two strip parts are as follows: With 
the shorthand notation $Q_{i,j,k,\cdots} = Q_i Q_j Q_k \cdots$, the left strip part, in Figure~\ref{fig:SO8Z2_vtx_LR}(\subref{fig:SO8Z2_vtx_La}), is given by
\begin{align}
& Z_{\mathrm{left}}^{\rm strip}(\mu_1, \mu_2, \mu_3, \nu_1, \nu_2) \nonumber \\
&= g^{\frac{1}{2}(\norm*{\mu_1}^2 + \norm*{\mu_2}^2 + \norm*{\mu_3}^2 + 2\norm*{\nu_1}^2 + 2\norm*{\nu_2^t}^2)} \tilde{Z}_{\mu_1}(g) \tilde{Z}_{\mu_2}(g) \tilde{Z}_{\mu_3}(g) \tilde{Z}_{\nu_1}(g)^2 \tilde{Z}_{\nu_2}(g)^2 \nonumber \\
&~~ \times \frac{\mathcal{R}_{\emptyset \emptyset}(Q_{1,2,3,4,9,10}) \mathcal{R}_{\emptyset \mu_1}(Q_{3,4,9,10}) \mathcal{R}_{\emptyset \mu_2}(Q_{3,4,9}) \mathcal{R}_{\emptyset \mu_3}(Q_{3,4}) \mathcal{R}_{\emptyset \nu_1}(Q_{1,1,2,3,4,9,10})  }
{\mathcal{R}_{\emptyset \mu_1^t}(Q_{1,2}) \mathcal{R}_{\emptyset \mu_2^t}(Q_{1,2,10}) \mathcal{R}_{\emptyset \mu_3^t}(Q_{1,2,9,10}) \mathcal{R}_{\emptyset \nu_1^t}(Q_{2,3,4,9,10}) \mathcal{R}_{\emptyset \nu_2}(Q_{1,2,3,3,4,9,10}) \mathcal{R}_{\mu_1 \mu_2^t}(Q_{10}) } \nonumber \\
&~~  \times \frac{\mathcal{R}_{\emptyset \nu_2^t}(Q_{1,2,4,9,10}) \mathcal{R}_{\mu_1^t \nu_1^t}(Q_2) \mathcal{R}_{\mu_2^t \nu_1^t}(Q_{2,10}) \mathcal{R}_{\mu_3^t \nu_1^t}(Q_{2,9,10}) \mathcal{R}_{\mu_1 \nu_2^t}(Q_{4,9,10}) \mathcal{R}_{\mu_2 \nu_2^t}(Q_{4,9}) }
{\mathcal{R}_{\mu_1 \mu_3^t}(Q_{9,10}) \mathcal{R}_{\mu_2 \mu_3^t}(Q_9) \mathcal{R}_{\mu_1^t \nu_1}(Q_{1,1,2}) \mathcal{R}_{\mu_2^t \nu_1}(Q_{1,1,2,10}) \mathcal{R}_{\mu_3^t \nu_1}(Q_{1,1,2,9,10}) \mathcal{R}_{\mu_1 \nu_2}(Q_{3,3,4,9,10})} \nonumber \\
&~~  \times\frac{\mathcal{R}_{\mu_3 \nu_2^t}(Q_4) \mathcal{R}_{\nu_1 \nu_1}(Q_{1,1}) \mathcal{R}_{\nu_2 \nu_2}(Q_{3,3}) \mathcal{R}_{\nu_1^t \nu_2}(Q_{2,3,3,4,9,10}) \mathcal{R}_{\nu_1 \nu_2^t}(Q_{1,1,2,4,9,10})}
{  \mathcal{R}_{\mu_2 \nu_2}(Q_{3,3,4,9}) \mathcal{R}_{\mu_3 \nu_2}(Q_{3,3,4}) \mathcal{R}_{\nu_1 \nu_2}(Q_{1,1,2,3,3,4,9,10})
    \mathcal{R}_{\nu_1^t \nu_2^t}(Q_{2,4,9,10})}\ ,
\end{align}
and the right strip part, in Figure~\ref{fig:SO8Z2_vtx_LR}(\subref{fig:SO8Z2_vtx_Lb}), is given by
\begin{align}
& Z_{\mathrm{right}}^{\rm strip}(\mu_1, \mu_2, \mu_3, \nu_3, \nu_4) \nonumber \\
&= g^{\frac{1}{2}(\norm*{\mu_1^t}^2 + \norm*{\mu_2^t}^2 + \norm*{\mu_3^t}^2 + 2\norm*{\nu_3^t}^2 + 2\norm*{\nu_4}^2)} \tilde{Z}_{\mu_1}(g) \tilde{Z}_{\mu_2}(g) \tilde{Z}_{\mu_3}(g) \tilde{Z}_{\nu_3}(g)^2 \tilde{Z}_{\nu_4}(g)^2 \nonumber \\
&~~\times \frac{\mathcal{R}_{\emptyset \emptyset}(Q_{5,6,7,8,9,10}) \mathcal{R}_{\emptyset \mu_1^t}(Q_{5,6}) \mathcal{R}_{\emptyset \mu_2^t}(Q_{5,6,10}) \mathcal{R}_{\emptyset \mu_3^t}(Q_{5,6,9,10}) \mathcal{R}_{\emptyset \nu_3^t}(Q_{6,7,8,9,10}) }{\mathcal{R}_{\emptyset \mu_1}(Q_{7,8,9,10}) \mathcal{R}_{\emptyset \mu_2}(Q_{7,8,9}) \mathcal{R}_{\emptyset \mu_3}(Q_{7,8}) \mathcal{R}_{\emptyset \nu_3}(Q_{5,5,6,7,8,9,10}) \mathcal{R}_{\emptyset \nu_4^t}(Q_{5,6,8,9,10}) \mathcal{R}_{\mu_1 \mu_2^t}(Q_{10})  } \nonumber \\
&~~\times \frac{ \mathcal{R}_{\emptyset \nu_4}(Q_{5,6,7,7,8,9,10}) \mathcal{R}_{\mu_1^t \nu_3^t}(Q_6) \mathcal{R}_{\mu_2^t \nu_3^t}(Q_{6,10}) \mathcal{R}_{\mu_3^t \nu_3^t}(Q_{6,9,10}) \mathcal{R}_{\mu_1 \nu_4^t}(Q_{8,9,10}) \mathcal{R}_{\mu_2 \nu_4^t}(Q_{8,9})  }{ \mathcal{R}_{\mu_1 \mu_3^t}(Q_{9,10}) \mathcal{R}_{\mu_2 \mu_3^t}(Q_9) \mathcal{R}_{\mu_1^t \nu_3}(Q_{5,5,6}) \mathcal{R}_{\mu_2^t \nu_3}(Q_{5,5,6,10}) \mathcal{R}_{\mu_3^t \nu_3}(Q_{5,5,6,9,10}) \mathcal{R}_{\mu_1 \nu_4}(Q_{7,7,8,9,10}) } \nonumber \\
&~~\times  \frac{\mathcal{R}_{\mu_3 \nu_4^t}(Q_8) \mathcal{R}_{\nu_3 \nu_3}(Q_{5,5}) \mathcal{R}_{\nu_4 \nu_4}(Q_{7,7}) \mathcal{R}_{\nu_3^t \nu_4}(Q_{6,7,7,8,9,10}) \mathcal{R}_{\nu_3 \nu_4^t}(Q_{5,5,6,8,9,10})  }{ \mathcal{R}_{\mu_2 \nu_4}(Q_{7,7,8,9}) \mathcal{R}_{\mu_3 \nu_4}(Q_{7,7,8}) \mathcal{R}_{\nu_3 \nu_4}(Q_{5,5,6,7,7,8,9,10}) \mathcal{R}_{\nu_3^t \nu_4^t}(Q_{6,8,9,10})}\ . 
\end{align}

As we are computing the partition function of a 6d theory on the $\Omega$-background which is compactified on a circle with a twist, it is a partition function on $\mathbb{R}^4\times T^2$.
Hence, it can be compared with the elliptic genus of 6d theories.
To this end, we expand the partition function \eqref{eq:SO8Z2_vtx} as 
\begin{align}
Z = Z_{\mathrm{pert}} \bigg(1 + \sum_{n=1}^{\infty} u^n Z_n \bigg)\ , 
\end{align}
where $Z_\mathrm{pert}$ is the perturbative part and $Z_n$ are the $n$-string elliptic genus, and $u$ is the string fugacity which is given by a product of K\"ahler parameters.

We proceed with the computation by expressing the K\"ahler parameters in terms of physical parameters. 
First we eliminate the dependence of $\phi_0$ along the W-bosons. For that, we perform the following shifts 
\begin{align}\label{eq:SO8Z2_varchange}
\phi_2 \to \phi_2 + 2\phi_0 + \frac{1}{2R}\ ,  \quad
\phi_3 \to \phi_3 + 2\phi_0 + \frac{1}{2R}\ , \quad
\phi_4 \to \phi_4 + \phi_0 + \frac{1}{4R}\ ,
\end{align}
which yield
\begin{align}\label{eq:SO8Z2_parameter}
&-\log \alpha_1 = 2\phi_4 - \phi_3\ , \quad \qquad
-\log \alpha_2 = 2\phi_3 - 2\phi_4 - \phi_2\ , \nonumber \\
&- \log \alpha_3 = 2\phi_2 - \phi_3\ , \quad \qquad
-\log \alpha_4 = -\phi_2 + \frac{\tau}{2}\ , \nonumber \\
&- \log Q_B = \phi_2 + \phi_3 + 4\phi_0 + \tau \ ,
\end{align}
where $\alpha_i$ and $Q_B$ are the K\"ahler parameters assigned in Figure~\ref{fig:SO8Z2_vtx}. Here, we have identified that $1/R$ as $\tau$, since it parametrizes the KK-momentum. One can see the Cartan matrix of $\mathfrak{so}(7)$ from the W-bosons in \eqref{eq:SO8Z2_parameter}, which is the invariant subalgebra under order two outer automorphism of $\mathfrak{so}(8)$ algebra. 
The string fugacity $u$ is $Q_B \alpha_1^{-2} \alpha_2^{-2} \alpha_3^{-1} \alpha_4$, and in terms of $u$, 
\begin{align}
- \log Q_B =- \log u + \phi_2 + \phi_3 - \frac{\tau}{2}\ .
\end{align}

As the perturbative part of the partition function $Z_{\mathrm{pert}}$ is the zeroth order in $u$, we obtain $Z_{\mathrm{pert}}$ by setting $\mu_i = \emptyset$ in \eqref{eq:SO8Z2_vtx}. It is convenient to  separate out the spinor matter parts, as  we will decouple them. To this end, we first sum over $\nu_i$ and express the K\"ahler parameters in terms of the $\alpha_i$ associated with the W-bosons, {\it e.g.}, $Q_2 = \alpha_4 Q_1^{-1}$, etc. Then we see that $Z_{\mathrm{pert}}$ is written as a function of $g$, $\alpha_i$ and $Q_1$, $Q_3$, $Q_5$, $Q_7$ that are the K\"ahler parameters associated with the spinor matter. 
Next, we take the plethystic logarithm, defined in Appendix~\ref{sec:special}, and expand  $Z_{\mathrm{pert}}$ in terms of $\alpha_i$ and $Q_i$. The shift \eqref{eq:SO8Z2_parameter} then yields
\begin{align}\label{eq:SO8Z2_pert_vtx}
Z_{\mathrm{pert}}
&= \operatorname{PE} \qty[\frac{2g}{(1-g)^2} \qty( \chi_{\Delta_+}^{\mathfrak{so}(7)} + (\chi_{\mathbf{7}}^{\mathfrak{so}(7)} - 1) q^{1/2} + \cdots )]\ ,
\end{align}
where $\operatorname{PE}$ stands for the plethystic exponential, 
\begin{align}
\operatorname{PE}[f(x)] = \exp(\sum_{n=1}^\infty \frac{1}{n} f(x^n))\ ,
\end{align}
and $q = e^{- \tau}$,  the terms $\cdots$ are the terms depend on the spinor matter.  Here
the characters for the positive roots  and the fundamental representation of  $\mathfrak{so}(7)$ algebra take the form
\begin{align}
\chi_{\Delta_+}^{\mathfrak{so}(7)}
&= x_2 + \frac{x_2^2}{x_3} + x_3 + \frac{x_3}{x_2} + \frac{x_2 x_3}{x_4^2} + \frac{x_3^2}{x_2 x_4^2} + \frac{x_4^2}{x_2} + \frac{x_4^2}{x_3} + \frac{x_2 x_4^2}{x_3}\ , \\
\chi_{\mathbf{7}}^{\mathfrak{so}(7)}
&= 1 + x_2 + \frac{1}{x_2} + \frac{x_2}{x_3} + \frac{x_3}{x_2} + \frac{x_3}{x_4^2} + \frac{x_4^2}{x_3}\ ,
\end{align}
where $x_i = e^{-\phi_i}$. 
We note that the $\cdots$ terms in \eqref{eq:SO8Z2_pert_vtx} denote the terms that involve $Q_1$, $Q_3$, $Q_5$ and $Q_7$ which we will decouple.  In the 5d limit where the KK-momentum becomes very large, the states with KK-momenta are truncated. Thus, only $q^0$ terms in $Z_{\mathrm{pert}}$ survive. 
It is then easy to see that that $Z_{\mathrm{pert}}$ involves the contribution from the positive roots of $\mathfrak{so}(7)$ which is expected on the Coulomb branch where masses of W-bosons are all positive.
This thus shows that the corresponding 5d theory is an $SO(7)$ gauge theory, as expected. 

The partition function for 6d self-dual strings can be expanded in terms of the string fugacity
\begin{align}\label{eq:Z_inst}
Z_{\mathrm{string}} = \frac{Z}{Z_{\mathrm{pert}}} = 1 + u Z_1 + u^2 Z_2 + \cdots\ ,
\end{align}
where $Z_n$ is the $n$-string elliptic genus. 
Again, having in mind the decoupling of the spinor contributions, when summing over Young diagrams  \eqref{eq:SO8Z2_vtx}, we express it as a function of $\alpha_i$ and $Q_1$, $Q_3$, $Q_5$, $Q_7$.  
To decouple the spinors, one needs to go through a flop transition, 
where masses of the spinors are proportional to $-\log$ of $Q_1^{-1}$, $Q_3^{-1}$, $Q_5^{-1}$, and $Q_7^{-1}$. 
We take each mass of the spinors to infinite or the corresponding K\"ahler parameters to zero while keeping $\alpha_1, \alpha_4$ finite.  
So, after decoupling the spinors, we get the desired $n$-string elliptic genus.

To organize $n$-string elliptic genus, we expand $Z_n$ in terms of the K\"ahler parameters associated with the W-bosons, given in \eqref{eq:SO8Z2_parameter}. In particular, using $\alpha_4= q^{1/2} (\alpha_1 \alpha_2 \alpha_3)^{-1}$, we expand  $Z_n$ in terms of $q$ and $\alpha_1$. 
The one-string elliptic genus $Z_1$ is expressed as 
\begin{small}
\begin{align}\label{eq:Z1 from 5-brane web}
Z_1&= \!\Bigg(\!-\frac{g \alpha_2^3 \alpha_3^2 \Big(\alpha_3 (\alpha_3+\!1) \alpha_2^2+\!(\alpha_3^2-6 \alpha_3+1) \alpha_2\!+\!\alpha_3\!+\!1\Big)}{(1-g)^2 (1-\alpha_2)^2 (1-\alpha_3)^2	(1-\alpha_2 \alpha_3)}\alpha_1^3 + O(\alpha_1^4)\Bigg) q^{-1/2} \nonumber\\
&~~ + \!\Bigg(-\frac{4g \alpha_2^2 \alpha_3^2 \Big((\alpha_3^2-\alpha_3+1) \alpha_2^2-(\alpha_3+1) \alpha_2+1\Big)}{(1-g)^2 (1-\alpha_2)^2 (1-\alpha_3)^2 (1-\alpha_2 \alpha_3)^2}\alpha_1^2 \nonumber\\
&\quad ~~ - \frac{4g \alpha_2^3 \alpha_3^2\Big (\alpha_3 (\alpha_3\!+\!1) \alpha_2^2+(\alpha_3^2\!-\!6 \alpha_3\!+\!1) \alpha_2\!+\!\alpha_3\!+\!1\Big)}{(1-g)^2 (1-\alpha_2)^2 (1-\alpha_3)^2 (1-\alpha_2 \alpha_3)^2} \alpha_1^3 +\! O(\alpha_1^4) \!\Bigg) +\! O(q^{1/2}). 
\end{align}
\end{small}
The two-string elliptic genus $Z_2$ is rather complicated, so we write only the leading terms in $\alpha_1$: 
\begin{small}
\begin{align}\label{eq:Z2 from 5-brane web}
Z_2&= \Bigg(\Bigg[\frac{g^5 \alpha_2^6 \alpha_3^4 
}{(1\!-\!g)^4 (1\!+\!g)^2 (1\!-\!\alpha_2)^2 (1\!-\!\alpha_3)^2}\qty(\frac{1}{(1\!-\!g \alpha_2)^2 (g-\alpha_3)^2} + 
\frac{1}{(g\!-\!\alpha_2)^2 (1-g \alpha_3)^2}
)  \nonumber\\
& +\! \frac{g^4 \alpha_2^6 \alpha_3^5}{(1-g)^4 (1-\alpha_2)^2 (1-\alpha_2 \alpha_3)^2 (g-\alpha_3)^2 (1-\alpha_3 g)^2} \nonumber \\
& +\! \frac{g^5 \alpha_2^6 \alpha_3^6 }{(1\!-\!g)^4 (1\!+\!g)^2 (1\!-\!\alpha_3)^2 (1\!-\!\alpha_2 \alpha_3)^2}\qty(\frac{1}{(g\!-\!\alpha_3)^2 (g\!-\!\alpha_2 \alpha_3)^2}\! +\! \frac{1}{(1\!-\!\alpha_3 g)^2 (1\!-\!\alpha_2 \alpha_3 g)^2}\!) \nonumber \\
& +\! \frac{g^4 \alpha_2^7 \alpha_3^4}{(1-g)^4 (g-\alpha _2)^2 (1-\alpha_2 g)^2 (1-\alpha_3)^2 (1-\alpha_2 \alpha_3)^2} \nonumber \\
&+\! \frac{g^4 \alpha_2^7 \alpha_3^5}{(1-g)^4 (1-\alpha_2)^2 (1-\alpha_3)^2 (g-\alpha_2 \alpha_3)^2 (1-\alpha_2 \alpha_3 g)^2} \nonumber \\
&+\! \frac{g^5 \alpha_2^8 \alpha_3^4 }{(1\!-\!g)^4 (1\!+\!g)^2 (1\!-\!\alpha_2)^2 (1\!-\!\alpha_2\alpha_3)^2}\!\qty(\!\frac{1}{(g\!-\!\alpha_2)^2 (g\!-\!\alpha _2 \alpha_3)^2}\! +\! \frac{1}{(1\!-\!g\alpha_2)^2 (1\!-\!g\alpha_2 \alpha_3)^2}\!) \!\Bigg]\! \alpha_1^6\crcr
& + O(\alpha_1^7)\Bigg) q^{-1} + O(q^{-1/2})\ .
\end{align}
\end{small}

In the following subsections, we will compare this obtained result with the elliptic genus computations in two different ways: First, as explained through 5-brane webs, we perform the Higgsing $SO(10)+2\mathbf{F}\to SO(9)+1\mathbf{F} \to SO(8)$ with $\mathbb{Z}_2$ twist. Secondly, we apply the $\mathbb{Z}$ twist directly to the elliptic genus of the 6d $SO(8)$ theory.

\subsection{Partition function from the Higgsing sequence} 

We now compute the partition function of the $SO(8)$ theory with $\mathbb{Z}_2$ twist  from direct Higgsing  of the partition function for the 6d $SO(10)$ gauge theory with two fundamental hypermultiplets. The perturbative part is given by
\begin{align}
Z_{\mathrm{pert}}^{\mathrm{gauge}}
= \operatorname{PE} \qty[ \frac{2g}{(1-g)^2} \qty( \chi_{\Delta_+}^{\mathfrak{so}(10)} + q \chi_{\Delta_-}^{\mathfrak{so}(10)}) \frac{1}{1-q} ]\ ,
\end{align}
where $\chi_{\Delta_+}^{\mathfrak{so}(10)}$ and $\chi_{\Delta_-}^{\mathfrak{so}(10)}$ are the positive and negative root parts of the  character for the adjoint representation of $\mathfrak{so}(10)$, respectively. More explicitly, for instance, $\chi_{\Delta_+}^{\mathfrak{so}(10)} = \sum_\alpha e^{- \alpha}$ for the positive roots $\alpha$, which can be also expressed in terms of 
$\phi_i$ in Figure~\ref{fig:SO10_flop1}(\subref{fig:SO10_flop1a}).

To realize the Higgsing $SO(10)+2F\to SO(9)+1F$, let us consider the contribution of the first hypermultiplet of mass $m_1$ to the perturbative part, which is given by 
\begin{align}
Z_{\mathrm{pert}}^{\mathrm{hyper1}} = \operatorname{PE} \qty[- \frac{g}{(1-g)^2} \qty( M_1 \chi_{\mathbf{10}}^{\mathfrak{so}(10)} + \frac{q}{M_1} \chi_{\mathbf{10}}^{\mathfrak{so}(10)}) \frac{1}{1-q} ]\ ,
\end{align}
where $\chi_{\mathbf{10}}^{\mathfrak{so}(10)}$ is the character for the fundamental representation $\mathbf{10}$ of $\mathfrak{so}(10)$, and $M_1 = e^{-m_1}$ is the fugacity for the mass of the fundamental hypermultiplet $m_1$. To Higgs the first hypermultiplet, we recover the affine node $\phi_0$ using Figure~\ref{fig:SO10_flop1}(\subref{fig:SO10_flop1a}) and impose \eqref{eq:SO8Z2_higgsing1}. If one changes the variables as in Figure~\ref{fig:6d_SO9}(\subref{fig:6d_SO9a}) and  
eliminate the affine node $\phi_0$, one can find that the perturbative part 
\begin{align}
Z_{\mathrm{pert}}^{\mathrm{gauge}} Z_{\mathrm{pert}}^{\mathrm{hyper1}}
\to \operatorname{PE}\qty[ \frac{2g}{(1-g)^2} \qty(\chi_{\Delta_+}^{\mathfrak{so}(9)} + q \chi_{\Delta_-}^{\mathfrak{so}(9)}) \frac{1}{1-q}  ]\ ,
\end{align}
up to the Cartan part, where $\chi_{\Delta_\pm}^{\mathfrak{so}(9)}$ are the positive and negative parts of the characters for the adjoint representation of $\mathfrak{so}(9)$, respectively. This reflects that the theory becomes the $SO(9)$ gauge theory as in Figure~\ref{fig:6d_SO9}(a).

Higgsing the second hypermultiplet of mass $m_2$ needs more caution because of $1/R$ factor in Figure~\ref{fig:6d_SO9}(b). From string tensions associated with the hypermultiplet in Figure~\ref{fig:6d_SO9}(b), one can read that the contribution of the second hypermultiplet after Higgsing the first hypermultiplet becomes
\begin{align}
Z_{\mathrm{pert}}^{\mathrm{hyper2}}
\to &\operatorname{PE}\bigg[  -\frac{g}{(1-g)^2} \bigg( M_2 \qty( \frac{x_1}{q} + \frac{x_1}{x_2} + \frac{x_2}{x_3} + \frac{x_3}{x_4^2} + 1 + \frac{x_4^2}{x_3} + \frac{x_3}{x_2} + \frac{x_2}{x_1} + \frac{q}{x_1} ) \nonumber \\
& + \frac{q}{M_2} \qty( \frac{x_1}{q} + \frac{x_1}{x_2} + \frac{x_2}{x_3} + \frac{x_3}{x_4^2} + 1 + \frac{x_4^2}{x_3} + \frac{x_3}{x_2} + \frac{x_2}{x_1} + \frac{q}{x_1}) \bigg) \frac{1}{1-q} \bigg]\ ,
\end{align}
where $x_i = e^{-\phi_i}$ for $\phi_i$ in Figure~\ref{fig:6d_SO9}(b). To Higgs the second hypermultiplet, we reintroduce the affine node $\phi_0$ again and impose \eqref{eq:SO8Z2_higgsing2}. Lastly, shifting variables \eqref{eq:SO8Z2_varchange} gives the perturbative part 
\begin{align}\label{eq:SO8Z2_pert_field}
Z_{\mathrm{pert}}^{\mathrm{gauge}} Z_{\mathrm{pert}}^{\mathrm{hyper1}} Z_{\mathrm{pert}}^{\mathrm{hyper2}}
\to \operatorname{PE} \qty[ \frac{2g}{(1-g)^2}\qty( (\chi_{\Delta_+}^{\mathfrak{so}(7)} - 2) + (\chi_{\mathbf{7}}^{\mathfrak{so}(7)} + 1) q^{1/2} + \cdots) \!],
\end{align}
which is the same result as  \eqref{eq:SO8Z2_pert_vtx} up to 
the Cartan part.

Next, consider the instanton part. The 
6d $SO(8+2p)$ gauge theories on $-4$ curve have 
$Sp(p)\times Sp(p)$ flavor symmetry.
In the 2d worldsheet theory on a self-dual string introduced in \cite{Haghighat:2014vxa}, both $SO(8+2p)$ gauge symmetry and $Sp(p)$ flavor symmetries become flavor symmetry. The gauge symmetry in 2d theory is $Sp(n)=USp(2n)$ where $n$ is string number. The field contents in this theory can be written as $\mathcal{N}=(0, 2)$ multiplets. The charged fields are vector multiplet $V$, three Fermi multiplets $\Lambda^\Phi, \Lambda^Q, \Lambda^{\tilde{Q}}$, and four chiral multiplets $B, \tilde{B}, Q, \tilde{Q}$. Their charges under the symmetries on a  worldsheet are summarized in Table~\ref{table:so(8+2p)}.
\begin{table}
\centering
\begin{tabular}{|c|c|c|c|c|c|c|} \hline
 & $Sp(n)$ & $SO(8+2p)$ & $Sp(p)_1$ & $Sp(p)_2$ & $U(1)_{\epsilon_1}$ & $U(1)_{\epsilon_2}$ \\ \hline
$V$ & $\mathbf{adj}$ & $\mathbf{1}$ & $\mathbf{1}$ & $\mathbf{1}$ & $0$ & $0$ \\ \hline
$\Lambda^\Phi$ & $\mathbf{adj}$ & $\mathbf{1}$ & $\mathbf{1}$ & $\mathbf{1}$ & $-1$ & $-1$ \\ \hline
$B$ & $\mathbf{antisymm}$ & $\mathbf{1}$ & $\mathbf{1}$ & $\mathbf{1}$ & $1$ & $0$ \\ \hline
$\tilde{B}$ & $\mathbf{antisymm}$ & $\mathbf{1}$ & $\mathbf{1}$ & $\mathbf{1}$ & $0$ & $1$ \\ \hline
$Q$ & $\mathbf{n}$ & $\mathbf{8+2p}$ & $\mathbf{1}$ & $\mathbf{1}$ & $1/2$ & $1/2$ \\ \hline
$\tilde{Q}$ & $\bar{\mathbf{n}}$ & $\overline{\mathbf{8+2p}}$ & $\mathbf{1}$ & $\mathbf{1}$ & $1/2$ & $1/2$ \\ \hline
$\Lambda^Q$ & $\mathbf{n}$ & $\mathbf{1}$ & $\mathbf{p}$ & $\mathbf{1}$ & $0$ & $0$ \\ \hline
$\Lambda^{\tilde{Q}}$ & $\bar{\mathbf{n}}$ & $\mathbf{1}$ & $\mathbf{1}$ & $\bar{\mathbf{p}}$ & $0$ & $0$ \\ \hline
\end{tabular}
\caption{The $\mathcal{N}=(0, 2)$ multiplets and symmetries on worldsheet in 6d $SO(8+2p)$ gauge theory.} \label{table:so(8+2p)}
\end{table}
The $U(1)_{\epsilon_1}$ and $U(1)_{\epsilon_2}$ are Cartans of $SO(4)$ rotating transverse $\mathbb{R}^4$ to worldsheet.

Using the field contents in Table~\ref{table:so(8+2p)}, one can calculate the elliptic genus of strings in the 6d $SO(10)$ theory using localization technique~\cite{Benini:2013xpa}. 
The $n$-string elliptic genus is expressed as
\begin{align}\label{eq:Zk_genus}
Z_n = \frac{1}{(2\pi i)^n} \frac{1}{\abs{\mathrm{Weyl}\qty[G]}} \oint Z_{\mathrm{1-loop}}\ ,
\end{align}
where $G$ is the gauge group on the 2d worldsheet. The contour integral is performed using the Jeffrey-Kirwan (JK) residue prescription in \cite{1993alg.geom..7001J}.

The one-loop determinant $Z_{\mathrm{1-loop}}$ is the product of the contributions from each multiplets. 
The $\mathcal{N}=(0, 2)$ vector multiplet contribution in $Z_{\mathrm{1-loop}}$ is given as a product of all the roots $\alpha$ of $G$ of rank $r$, 
\begin{align}
Z_{\mathrm{vec}} = \qty(\frac{2\pi \eta^2}{i})^r \prod_{\alpha} \frac{i \theta_1(\alpha)}{\eta} \prod_{j=1}^r du_j\ ,
\end{align}
where $\eta$ is the Dedekind eta function, $\theta_1(x)$ is the Jacobi theta function. See Appendix~\ref{sec:special} for the definition and some useful properties of the theta function.  
The $\mathcal{N}=(0, 2)$ chiral and Fermi multiplet contributions are given as
\begin{align}
Z_{\mathrm{chiral}} = \prod_\rho \frac{i \eta}{\theta_1(\rho)}, \qquad Z_{\mathrm{Fermi}} = \prod_\rho \frac{i \theta_1(\rho)}{\eta}\ ,
\end{align}
where $\rho$ is the weight vector of the representation of each multiplet in gauge and flavor symmetries.

In the case of the $SO(10)$ theory, the 2d gauge group is $Sp(n)$. Note that in orthogonal basis $\{e_i\}_{i=1}^n$ of $\mathfrak{sp}(n)$ algebra, the root vectors are given by $\pm e_i \pm e_j$, the weight vectors of the fundamental and the antisymmetric representations are $\pm e_i$ and $\pm e_i \pm e_j(i \neq j$), respectively. Denote chemical potential parameters of $\mathfrak{sp}(n)$ gauge algebra be $u_i$ and $\mathfrak{so}(10)$, $\mathfrak{sp}(1)_1$, $\mathfrak{sp}(1)_2$ flavor symmetry algebras be $v_i$, $m_1$ and $m_2$, respectively. Then the $n$-string elliptic genus from  \eqref{eq:Zk_genus} is
\begin{align}\label{eq:SO10_genus}
Z_n^{SO(10)}
&= \frac{1}{(2\pi i)^n } \frac{1}{\abs{\mathrm{Weyl}[Sp(n)]}} \oint \prod_{I=1}^n du_I \cdot \qty(\prod_{I=1}^n \frac{i^2 \theta_1(\pm 2 u_I)}{\eta^2}) \qty(\prod_{I<J}^n \frac{i^4 \theta_1(\pm u_I \pm u_J)}{\eta^4}) \nonumber \\
& \quad\times \qty(\frac{i \theta_1(-2\epsilon_+)}{\eta})^n \qty(\prod_{I=1}^n \frac{i^2 \theta_1(-2\epsilon_+ \pm 2u_I)}{\eta^2}) \qty(\prod_{I<J}^n \frac{i^4 \theta_1(-2\epsilon_+ \pm u_I \pm u_J)}{\eta^4}) \nonumber \\
& \quad \times\qty(\frac{i^2 \eta^2}{\theta_1(\epsilon_{1,2})})^n \qty(\prod_{I<J}^n \frac{i^8 \eta^8}{\theta_1(\epsilon_{1,2} \pm u_I \pm u_J)}) \qty(\prod_{I=1}^n \prod_{J=1}^5 \frac{i^4 \eta^4}{\theta_1(\epsilon_+ \pm u_I \pm v_J)}) \nonumber \\
& \quad \times\qty(\prod_{I=1}^n \frac{i^4 \theta_1(\pm u_I + m_1) \theta_1(\pm u_I + m_2)}{\eta^4})\ ,
%
\end{align}
where $\epsilon_\pm = \frac{\epsilon_1 \pm \epsilon_2}{2}$ and we use shorthand notation $\theta_1(\pm 2u_i) = \theta_1(2u_i) \theta_1(-2u_i)$, $\theta_1(\epsilon_{1,2}) = \theta_1(\epsilon_1) \theta_1(\epsilon_2)$, etc.\footnote{For example, $\theta_1(\epsilon_{1,2} \pm u_I \pm u_J)=\!\displaystyle\prod_{n=1}^{2}\!\theta_1(\epsilon_{n} \!+ u_I + u_J)\theta_1(\epsilon_{n}\! + u_I - u_J)\theta_1(\epsilon_{n} \!- u_I + u_J)\theta_1(\epsilon_{n} \!-u_I - u_J)$.}

To evaluate the integral \eqref{eq:SO10_genus}, one needs to find residues of the integrand. However, not all poles contributes to elliptic genus. The contributing residues are called Jeffrey-Kirwan (JK) residue, determined as follows. Suppose the singular hyperplanes $H_n$ are $\sum_{I=1}^r a_I^{(n)} u_I + \cdots = 0$. 
Fix an auxiliary vector $\zeta$ in the space which contains vectors $a^{(n)}$. If $\zeta$ can be written as linear combination of $a^{(n_1)}, \cdots , a^{(n_l)}$ with positive coefficients, then the singularities $H_1, \cdots, H_l$ contribute to the integral.

\begin{figure}
\centering
\includegraphics[scale=1]{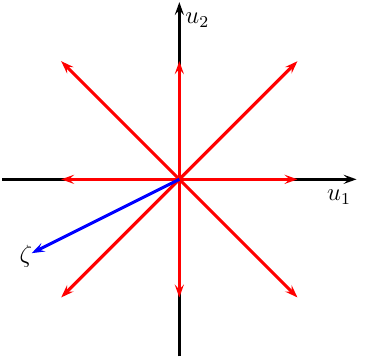}
\caption{A choice of an auxiliary vector $\zeta$ for JK-residue calculation of two-string elliptic genus.} \label{fig:SO10_JK}
\end{figure}

We explicitly calculate the one string and two string elliptic genus to compare with topological vertex calculation. For $n=1$, the singular hyperplanes are
\begin{align}
\pm u_1 \pm v_J + \epsilon_+ = 0\ .
\end{align}
If one takes an auxiliary vector $\zeta=(1)$, only singular hyperplanes $+u_1 \pm v_J + \epsilon_+ = 0$ 
contribute to the elliptic genus. Using
\begin{align}
\frac{1}{2\pi i} \oint_{u_1=0}  \frac{du}{\theta_1(u_1)} = \frac{1}{2\pi \eta^3}\ , \qquad \theta_1(-x) = -\theta_1(x)\ ,
\end{align}
the one-string elliptic genus is
\begin{align}\label{eq:SO10_Z1}
Z_1^{SO(10)}
&= -\frac{\eta^{12}}{2} \sum_{I=1}^5 \Bigg(\frac{\theta_1(2\epsilon_+ + 2v_I) \theta_1(4\epsilon_+ + 2v_I) \theta_1(\epsilon_+ + v_I \pm m_1) \theta_1(\epsilon_+ + v_I \pm m_2)}{\theta_1(\epsilon_{1,2}) \prod_{J\neq I}^5  \theta_1(v_I \pm v_J) \theta_1(2\epsilon_+ + v_I \pm v_J)}  \nonumber \\
& \qquad \qquad \qquad \qquad + (v_I \to -v_I) \Bigg).
%
%
\end{align}
For $n=2$, the singular hyperplanes are
\begin{align}
\pm u_I \pm u_2 + \epsilon_{1,2} = 0, \qquad \pm u_I \pm v_J + \epsilon_+ = 0\ , 
\end{align}
for $I=1,2$ and $J=1, \cdots, 5$. 
We choose an auxiliary vector $\zeta$ as in Figure~\ref{fig:SO10_JK}, so that the contributing poles are
\begin{align}\label{eq:SO(10)_pole}
(i) \left\{
\begin{array}{l}
-u_1 \pm v_I + \epsilon_+ = 0 \\
-u_1 - u_2 + \epsilon_{1,2} = 0
\end{array}\right.
& &&
(ii) \left\{
\begin{array}{l}
-u_1 \pm v_I + \epsilon_+ = 0  \\
u_1 - u_2 + \epsilon_{1,2} = 0
\end{array}\right. \nonumber \\
(iii) \left\{
\begin{array}{l}
-u_1 \pm v_I + \epsilon_+ = 0 \\
-u_2 \pm v_J + \epsilon_+ = 0 \\
\end{array}\right.
& &&
(iv) \left\{
\begin{array}{l}
-u_1 + u_2 + \epsilon_{1,2} = 0 \\
-u_1 - u_2 + \epsilon_{1,2} = 0
\end{array}\right. \\
(v) \left\{
\begin{array}{l}
-u_1 + u_2 + \epsilon_{1,2} = 0 \\
-u_2 \pm v_I + \epsilon_+ = 0 \\
\end{array}\right.
& &&
(vi) \left\{
\begin{array}{l}
u_2 \pm v_I + \epsilon_+ = 0 \\
-u_1 - u_2 + \epsilon_{1,2} = 0
\end{array}\right.\ . \nonumber 
\end{align}
The residue around poles \eqref{eq:SO(10)_pole}(\textit{i}) is
\begin{align}\label{eq:SO10_genus_pole1}
\mathrm{Res}_1
&= \frac{\eta^{24}}{8} \sum_{I=1}^5 \bigg[ \frac{\theta_1(3\epsilon_+ + \epsilon_-) \theta_1(2\epsilon_- - 2v_I) \theta_1(2\epsilon_1 - 2v_I) \theta_1(4\epsilon_+ + 2v_I) \theta_1(\epsilon_2 + 2v_I)}{\theta_1(\epsilon_1) \theta_1(\epsilon_2)^2 \theta_1(2\epsilon_1) \theta_1(2\epsilon_-) \theta_1(2v_I)} \nonumber \\
& \qquad \quad \times\!\! \frac{ \theta_1(3\epsilon_+ - \epsilon_- + 2v_I) \theta_1(\epsilon_+ + v_I \pm m_{1,2}) \theta_1(\epsilon_- - v_I \pm m_{1,2})}{ \prod_{J \neq I}^5\theta_1(v_I \pm v_J) \theta_1(\epsilon_1 - v_I \pm v_J)\theta_1(\epsilon_2 + v_I \pm v_J) \theta_1(2\epsilon_+ + v_I \pm v_J)} \nonumber \\
& \qquad \qquad  + (v_I \to -v_I) \bigg] + (\epsilon_1 \leftrightarrow \epsilon_2, \: \epsilon_- \to -\epsilon_-)\ .
\end{align}
The poles \eqref{eq:SO(10)_pole}(\textit{ii}) give
\begin{align}\label{eq:SO10_genus_pole2}
\mathrm{Res}_{2}
&= -\frac{\eta^{24}}{8} \sum_{I=1}^5 \bigg[ \frac{\theta_1(3\epsilon_+ + \epsilon_- + 2v_I) \theta_1(4\epsilon_+ + 2\epsilon_- + 2v_I) \theta_1(5\epsilon_+ + \epsilon_- + 2v_I)}{\theta_1(\epsilon_{1,2}) \theta_1(2\epsilon_1) \theta_1(2\epsilon_-)} \nonumber \\
& \qquad \quad \times \!\! \frac{\theta_1(6\epsilon_+ \!+ 2\epsilon_- + 2v_I) \theta_1(\epsilon_+\! + v_I \pm m_{1,2}) \theta_1(2\epsilon_+ + \epsilon_-\! + v_I \pm m_{1,2})}{\prod_{J\neq I}^5\theta_1(v_I \pm v_J) \theta_1(\epsilon_1\! + v_I \pm v_J) \theta_1(2\epsilon_+\! + v_I \pm v_J) \theta_1(3\epsilon_+ \!+ \epsilon_-\! + v_I \pm v_J)} \nonumber \\
& \qquad \qquad + (v_I \to -v_I) \bigg] + (\epsilon_1 \leftrightarrow \epsilon_2, \: \epsilon_- \to -\epsilon_-)\ ,
\end{align}
and the poles \eqref{eq:SO(10)_pole}(\textit{iii}) contribute
\begin{align}\label{eq:SO10_genus_pole3}
\mathrm{Res}_{3}
&= \frac{\eta^{24}}{8} \sum_{I \neq J}^5 \bigg[ \frac{\theta_1(2\epsilon_+ + 2v_{I,J}) \theta_1(4\epsilon_+ + 2v_{I,J}) \theta_1(4\epsilon_+ + v_I + v_J)}{\theta_1(\epsilon_{1,2})^2 \theta_1(v_I + v_J) \theta_1(\epsilon_{1,2} + v_I + v_J) \theta_1(\epsilon_{1,2} \pm (v_I - v_J))} \nonumber \\
& \qquad \quad \times\!\! \frac{\theta_1(\epsilon_+ + v_{I,J} \pm m_{1,2})}{ \theta_1(3\epsilon_+ \pm \epsilon_- + v_I + v_J) \prod_{K\neq I, J}^5 \theta_1(v_{I,J} \pm v_K) \theta_1(2\epsilon_+ + v_{I,J} \pm v_K)} \nonumber \\
& \quad \qquad + (v_I \to -v_I) + (v_J \to -v_J) + (v_I \to -v_I, \: v_J \to -v_J) \bigg],
\end{align}
where $\theta_1(2\epsilon_+ + 2v_{I,J}) =\theta_1(2\epsilon_+ + 2v_{I})\theta_1(2\epsilon_+ + 2v_{J})$.  
The residue of poles \eqref{eq:SO(10)_pole}(\textit{iv}) is zero. The cases \eqref{eq:SO(10)_pole}(\textit{v}) and (\textit{vi}) are $u_1 \leftrightarrow u_2$ and $u_1 \to u_2, \: u_2 \to -u_1$ of the case ($\textit{ii}$), respectively. These transformations do not change the residues. As a result, two string elliptic genus reads
\begin{align}\label{eq:SO10_Z2}
Z_2^{SO(10)} = \mathrm{Res}_1+ 3 \times \mathrm{Res}_2 + \mathrm{Res}_3\ .
\end{align}

We now Higgs this $SO(10)$ elliptic genera. We only consider the unrefined case, $\epsilon_1 = -\epsilon_2$. The brane web construction gives hints for Higgsing procedure. In the elliptic genus calculated from the field theory construction, there are $\mathfrak{so}(10)$ parameters $v_i$. They are written in the orthogonal basis, so first convert it to fundamental weight basis (Dynkin basis) and recover an affine node:
\begin{align}
& v_1 = \phi_1 - \phi_0, \qquad
v_2 = -\phi_0 - \phi_1 + \phi_2, \nonumber \\
& v_3 = \phi_3 - \phi_2, \qquad
v_4 = -\phi_3 + \phi_4 + \phi_5, \qquad
v_5 = \phi_5 - \phi_4
\end{align}
Then we can apply the Higgsing conditions \eqref{eq:SO8Z2_higgsing1} and \eqref{eq:SO8Z2_higgsing2}. Lastly, we shift variables \eqref{eq:SO8Z2_varchange} to compare with topological vertex calculation. The result is
\begin{align}
v_1 = 0, \quad
v_2 = -2\phi_4 + \phi_3, \quad
v_3 = \phi_2 - \phi_3, \quad
v_4 = -\phi_2, \quad
v_5 = \frac{\tau}{2}
\end{align}
Note that $v_2$, $v_3$, $v_4$ form the orthogonal basis of $\mathfrak{so}(7)$ algebra. Thus, by redefinition of the variables, the Higgsing sets the parameters as
\begin{align}\label{eq:SO8Z2_higgsing_genus}
m_1 = 0\ , \qquad m_2 = \frac{\tau}{2}\ , \qquad v_4 = \frac{\tau}{2}\ , \qquad v_5 = 0\ .
\end{align}

Substituting \eqref{eq:SO8Z2_higgsing_genus} into \eqref{eq:SO10_Z1} and \eqref{eq:SO10_Z2} and taking the unrefined limit $\epsilon_1 = -\epsilon_2$, we find that the one-string elliptic genus is given by
\begin{align}\label{eq:SO8Z2_Z1_genus}
Z_1 =\! \frac{\eta^{12}}{2 \theta_1(\epsilon_-)^2} \sum_{I=1}^3 \!\qty( \frac{\theta_1(2v_I)^2}{\theta_1(v_I)^2 \theta_1(v_I\! -\! \frac{\tau}{2}) \theta_1(v_I\! +\! \frac{\tau}{2}) \prod_{J\neq I}^3 \theta_1(v_I \pm v_J)^2} + (v_I\!\to\! -v_I)) ,
\end{align}
and the two-string elliptic genus is given by
\begin{align}\label{eq:SO8Z2_Z2_genus}
Z_2 = \mathrm{Res}_1 + 3 \times \mathrm{Res}_2 + \mathrm{Res}_3\ ,
\end{align}
where 
\begin{align}
\mathrm{Res}_1
&= \frac{\eta^{24}}{8}\sum_{I=1}^3 \bigg[ \frac{ \theta_1(\epsilon_- - 2v_I)^2 \theta_1(2\epsilon_- - 2v_I)^2}{\theta_1(\epsilon_-)^2 \theta_1(2\epsilon_-)^2 \theta_1(v_I)^2 \theta_1(\epsilon_- - v_I)^2 \theta_1(\pm v_I + \frac{\tau}{2}) \theta_1(\pm (\epsilon_- - v_I) + \frac{\tau}{2})} \nonumber \\
& \qquad \qquad\times \prod_{J \neq I}^3\frac{1}{ \theta_1(\epsilon_- - v_I - v_J)^2 \theta_1(v_I \pm v_J)^2 \theta_1(\epsilon_- - v_I + v_J)^2 }  + (v_I \to -v_I) \bigg] \nonumber \\
& \qquad  + (
\epsilon_- \to -\epsilon_-)\ ,\\
\mathrm{Res}_2
&= \frac{\eta^{24}}{8} \sum_{I=1}^3 \bigg[ \frac{\theta_1(\epsilon_- + 2v_I)^2 \theta_1(2\epsilon_- + 2v_I)^2}{\theta_1(\epsilon_-)^2 \theta_1(2\epsilon_-)^2 \theta_1(v_I)^2 \theta_1(\epsilon_- + v_I)^2 \theta_1(\pm v_I + \frac{\tau}{2}) \theta_1(\pm (\epsilon_- + v_I) + \frac{\tau}{2})} \nonumber \\
& \qquad \qquad \times\prod_{J \neq I}^3\frac{1}{ \theta_1(v_I \pm v_J)^2 \theta_1(\epsilon_- + v_I \pm v_J)^2} + (v_I \to -v_I) \bigg] \nonumber \\
& \quad  + ( 
\epsilon_- \to -\epsilon_-)\ ,\\
\mathrm{Res}_3
&= \frac{\eta^{24}}{8} \sum_{I \neq J}^3 \bigg[ \frac{\theta_1(2v_{I,J})^2}{\theta_1(\epsilon_-)^4 \theta_1(v_{I,J})^2 \theta_1(\epsilon_- \pm v_I \pm v_J)^2 \theta_1(\pm v_{I,J} + \frac{\tau}{2}) 
\prod_{K \neq I, J}^3\theta_1(v_{I,J} \pm v_K)^2} \nonumber \\
& \qquad \qquad + (v_I \to -v_I) + (v_J \to -v_J) + (v_I \to -v_I, \: v_J \to -v_J) \bigg].
\end{align}

Since $v_i$ are the $\mathfrak{so}(7)$ parameters, we change the variables
\begin{align}
v_1 &= - \log \alpha_1\ , \qquad
v_2 = -\qty( \log \alpha_1 + \log \alpha_2)\ , \cr
v_3 &= - \qty( \log \alpha_1 + \log \alpha_2 + \log \alpha_3)\ .
\end{align}
and expand $Z_1$ and $Z_2$ in terms of $q$ and $\alpha_1$ to compare with topological vertex calculation in the previous subsection. By an explicit calculation, we see that two results are in complete agreement. 

\subsection{Partition function from a direct twisting of the \texorpdfstring{$SO(8)$}{SO8} theory}\label{sec:directTwistingSO8}

There is yet another way to obtaining the partition function of the $SO(8)$ theory with $\mathbb{Z}_2$ twist from the field theory construction. The perturbative part of the partition function of the 6d $SO(8)$ theory is given by
\begin{align}\label{eq:SO8_pert}
Z_{\mathrm{pert}}^{SO(8)}
= \operatorname{PE}\qty[ \frac{2g}{(1-g)^2} \qty(\chi_{\Delta_+}^{\mathfrak{so}(8)} + q \chi_{\Delta_-}^{\mathfrak{so}(8)}) \frac{1}{1-q} ]\ ,
\end{align}
where 
$\chi_{\Delta_\pm}^{\mathfrak{so}(8)}$ are the positive and negative parts of the character for the adjoint representation of $\mathfrak{so}(8)$,
\begin{align}
\chi_{\Delta_+}^{\mathfrak{so}(8)}
= \sum_{i<j}^4 \qty(x_i x_j + \frac{x_i}{x_j}), \qquad
\chi_{\Delta_-}^{\mathfrak{so}(8)}
= \sum_{i<j}^4 \qty(\frac{1}{x_i x_j} + \frac{x_j}{x_i})\ ,
\end{align}
for the fugacities $x_i$ of the orthonormal basis of $\mathfrak{so}(8)$.

The elliptic genus of the 6d pure $SO(8)$ theory can be obtained from the 2d worldsheet theory with matters in Table~\ref{table:so(8+2p)}, with $p=0$. 
As there is no Fermi multiplet $\Lambda^Q$ and $\Lambda^{\tilde{Q}}$ for $p=0$, 
the $n$-string elliptic genus is given by 
\begin{align}\label{eq:SO8_genus}
Z_n^{SO(8)}
&= \frac{1}{(2\pi i)^n } \frac{1}{\abs{\mathrm{Weyl}[Sp(n)]}} \oint \prod_{I=1}^n du_I \cdot \qty(\prod_{I=1}^n \frac{i^2 \theta_1(\pm 2 u_I)}{\eta^2}) \qty(\prod_{I<J}^n \frac{i^4 \theta_1(\pm u_I \pm u_J)}{\eta^4}) \nonumber \\
& \quad \times \qty(\frac{i \theta_1(-2\epsilon_+)}{\eta})^n \qty(\prod_{I=1}^n \frac{i^2 \theta_1(-2\epsilon_+ \pm 2u_I)}{\eta^2}) \qty(\prod_{I<J}^n \frac{i^4 \theta_1(-2\epsilon_+ \pm u_I \pm u_J)}{\eta^4}) \nonumber \\
& \quad \times \qty(\frac{i^2 \eta^2}{\theta_1(\epsilon_{1,2})})^n \qty(\prod_{I<J}^n \frac{i^8 \eta^8}{\theta_1(\epsilon_{1,2} \pm u_I \pm u_J)}) \qty(\prod_{I=1}^n \prod_{j=1}^4 \frac{i^4 \eta^4}{\theta_1(\epsilon_+ \pm u_I \pm v_J)}) .
\end{align}
$v_J$ are the chemical potentials for the  $SO(8)$ symmetry and they can be regarded as being in the orthogonal basis of $SO(8)$.

Let the eight states in the fundamental representation be $\ket{\pm e_i}$ where $i=1, 2, 3, 4$. They satisfy $H^j \ket{\pm e_j} = \pm \delta_{ij} \ket{\pm e_j}$ for the orthonormal basis $H^j$ of the Cartan subalgebra of $\mathfrak{so}(8)$.
The order two outer automorphism $\sigma$ of $\mathfrak{so}(8)$ algebra exchanges the simple root $\alpha_3$ and $\alpha_4$. In terms of the orthonormal basis of the Cartan subalgebra, $\sigma(H^4) = -H^4$. Thus, the invariant elements are $\pm H^1$, $\pm H^2$, $\pm H^3$ and $H^4 - H^4$, corresponding to the states $\ket{\pm e_1}$, $\ket{\pm e_2}$, $\ket{\pm e_3}$ and $\ket{e_4} + \ket{-e_4}$. On the other hand, $H^4 - (-H^4)$ has the eigenvalue $-1$, which corresponds to the state $\ket{e_4} - \ket{-e_4}$. Upon compactification, $\ket{e_4}$ and $\ket{-e_4}$ are identified so that the invariant states $\ket{\pm e_1}$, $\ket{\pm e_2}$, $\ket{\pm e_3}$ and $\ket{0}$ form the fundamental representation of the invariant algebra $\mathfrak{so}(7)$ under the order two outer automorphism. The state of eigenvalue $-1$ becomes a singlet and gets additional half KK-momentum shift~\cite{Tachikawa:2011ch}.

Consequently, we change $(x_4 + \frac{1}{x_4}) \sum_{i=1}^3 x_i$ to $(1+q^{1/2}) \sum_{i=1}^3 x_i$ in $\chi_{\Delta_+}^{\mathfrak{so}(8)}$
and $(v_4 + \frac{1}{v_4}) \sum_{i=1}^3 \frac{1}{v_i}$ to $(1+q^{-1/2}) \sum_{i=1}^3 \frac{1}{v_i}$ in 
$\chi_{\Delta_-}^{\mathfrak{so}(8)}$. After this replacement, the perturbative part becomes 
\begin{align}
Z_{\mathrm{pert}}
= \operatorname{PE}\qty[ \frac{2g}{(1-g)^2} \qty( \chi_{\Delta_+}^{\mathfrak{so}(7)} + \qty(\chi_{\mathbf{7}}^{\mathfrak{so}(7)}-1) q^{1/2} + \cdots )]\ ,
\end{align}
which agrees with the perturbative part obtained from 5-brane webs in \eqref{eq:SO8Z2_pert_vtx}.
For the instanton partition function, we replace $\pm v_4$ in \eqref{eq:SO8_genus} to $\frac{\tau}{2}$ and $0$. The elliptic genus is then expressed as 
\begin{align}
Z_n
&= \frac{1}{(2\pi i)^n } \frac{1}{\abs{\mathrm{Weyl}[Sp(n)]}} \oint \prod_{I=1}^n du_I \cdot \qty(\prod_{I=1}^n \frac{i^2 \theta_1(\pm 2 u_I)}{\eta^2}) \qty(\prod_{I<J}^n \frac{i^4 \theta_1(\pm u_I \pm u_J)}{\eta^4}) \nonumber \\
& \quad \times \qty(\frac{i \theta_1(-2\epsilon_+)}{\eta})^n \qty(\prod_{I=1}^n \frac{i^2 \theta_1(-2\epsilon_+ \pm 2u_I)}{\eta^2}) \qty(\prod_{I<J}^n \frac{i^4 \theta_1(-2\epsilon_+ \pm u_I \pm u_J)}{\eta^4}) \nonumber \\
& \quad \times \qty(\frac{i^2 \eta^2}{\theta_1(\epsilon_{1,2})})^n \qty(\prod_{I<J}^n \frac{i^8 \eta^8}{\theta_1(\epsilon_{1,2} \pm u_I \pm u_J)}) \qty(\prod_{I=1}^n \prod_{J=1}^3 \frac{i^4 \eta^3}{\theta_1(\epsilon_+ \pm u_I \pm v_J)}) \nonumber \\
& \quad \times\prod_{I=1}^n \frac{i^4 \eta^3}{\theta_1(\epsilon_+ \pm u_I) \theta_1(\epsilon_+ \pm u_I + \frac{\tau}{2})}\ .
\end{align}
For $n=1$, the singular hyperplanes are
\begin{align}
\pm u_1 + \epsilon_+ \pm v_J = 0\ , \qquad \pm u_1 + \epsilon_+ = 0\ , \qquad \pm u_1 + \epsilon_+ + \frac{\tau}{2} = 0\ .
\end{align}
Take an auxiliary vector $\zeta=(1)$. The residues of the last two singular hyperplanes are zero since $\theta_1(k \tau) = 0$ for $k \in \mathbb{Z}$. Thus, only poles $+u_1 + \epsilon_+ \pm v_j=0$ contribute to the one string elliptic genus. One can find that the JK-residue sum for $Z_1$ with the unrefined limit $\epsilon_1 = -\epsilon_2$ gives the same expression as \eqref{eq:SO8Z2_Z1_genus}. A similar calculation for the $n=2$ case leads to \eqref{eq:SO8Z2_Z2_genus} as well.

\section{\texorpdfstring{$SU(3)$}{SU3} theory with \texorpdfstring{$\mathbb{Z}_2$}{Z2} twist} \label{sec:SU3}

\begin{figure}
\centering
\includegraphics[scale=1]{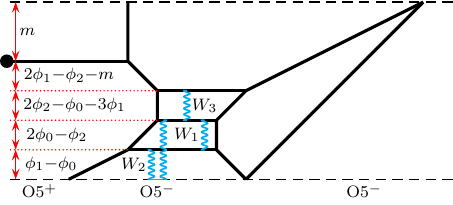}
\caption{A 5-brane web for 6d $G_2$ gauge theory with one fundamental.  } 
 \label{fig:G2}
\end{figure}

The 6d $SU(3)$ gauge theory with $\mathbb{Z}_2$ twist can be obtained by Higgsing from the 6d $G_2$ gauge theory with a hypermultiplet in the fundamental representation. The Higgsing procedure can be realized through 5-brane webs as given in~\cite{Kim:2019dqn}. It is also worthy of noting that from the perspective of 5d SCFTs, it is 5d pure $SU(3)$ gauge theory at the Chern-Simons level $\kappa=9$ ~\cite{Jefferson:2018irk}. The corresponding 5-brane web is given in~\cite{Hayashi:2018lyv}. In order not to be confused with this 5d $SU(3)$ theory, we denote it  by $SU(3)_9$. In this section, we compute the partition function for the 6d $SU(3)$ gauge theory with $\mathbb{Z}_2$ twist from 5-brane webs using topological vertex and also from performing a Higgsing on the elliptic genera of self-dual strings in the 6d $G_2$ theory. 


\subsection{5-brane web for 6d \texorpdfstring{$SU(3)$}{SU3} gauge theory with \texorpdfstring{$\mathbb{Z}_2$}{Z2}
twist}

We first review Type IIB 5-brane construction of the 6d pure $SU(3)$ gauge theory with $\mathbb{Z}_2$ twist. 
We start from the 6d $G_2$ gauge theory with one fundamental hypermultiplet, whose 5-brane configuration is given in Figure~\ref{fig:G2}\footnote{The 5-brane web given in Figure~\ref{fig:G2} is obtained by taking monodromy cut of D7-branes on O5-planes differently compared with the 5-brane web given in \cite{Kim:2019dqn}. One can think of it as an $SL(2,\mathbb{Z})$ transformed web diagram.}, where the mass of the fundamental hyper is denoted by $m$ and the masses of W-bosons $W_i$ are expressed in terms of the scalar fields as
\begin{align}
m_{W_1} = 2\phi_0 - \phi_2\ , \qquad
m_{W_2} = 2\phi_1 - \phi_2\ , \qquad
m_{W_3} = -\phi_0 - 3\phi_1 + 2\phi_2\ ,
\end{align}
which form the affine Cartan matrix of untwisted affine Lie algebra $G_2^{(1)}$. 

\begin{figure}
\centering
\includegraphics[scale=0.9]{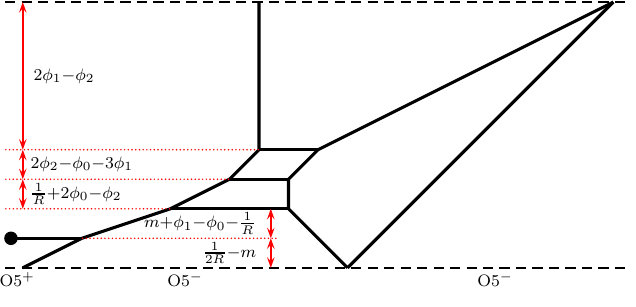}
\caption{Lowering the flavor brane in Figure~\ref{fig:G2} through a series of flop transitions, so as to apply the Higgsing procedure discussed in Figure~\ref{fig:SO10_flop1}(b)-(d). } \label{fig:G2_flop}
\end{figure}

To obtain the $SU(3)$ gauge theory with $\mathbb{Z}_2$ twist from the $G_2$ gauge theory with a fundamental, we Higgs the hypermultiplet of the $G_2$ gauge theory by giving vevs to the scalar fields which carry KK-momentum. To this end, 
we redefine the scalars as
\begin{align}
2\phi_0 - \phi_2 \to 2\phi_0 - \phi_2 + \frac{1}{R}\ , \qquad
\phi_1 - \phi_0 \to \phi_1 - \phi_0 - \frac{1}{2R}\ ,
\end{align}
while other parameters in Figure~\ref{fig:G2} remain unaltered. This recovers the 6d circle radius $R$ in the brane configuration. 
The Higgsing procedure to the 6d $SU(3)$ gauge theory with $\mathbb{Z}_2$ twist is then similar to the Higgsing explained through Figure~\ref{fig:SO10_flop1}(b)-(d) in the previous section. 
We first bring the flavor D7-brane down to the lower O5-plane as depicted in Figure~\ref{fig:G2_flop}. The Higgsing condition is then given by
\begin{align}\label{eq:G2_higgsing}
m = \frac{1}{2R}, \qquad \phi_1 = \phi_0 + \frac{1}{2R}\ .
\end{align}

The first condition in \eqref{eq:G2_higgsing} comes from the flavor D7-brane put on the lower $\mathrm{O}5$-plane. As discussed in the previous section, this D7-brane is brought to the lower chamber of the Coulomb branch and then is put to the $\mathrm{O}5^-$-plane on the bottom. It follows that this full D7-brane is split into two half D7-branes generating an $\widetilde{\mathrm{O}5}^-$-plane in between these two half D7-branes. The second condition in \eqref{eq:G2_higgsing} comes from putting the lower color D5-brane on the $\mathrm{O}5^-$-plane in the middle, which also becomes two half D5-branes. Through Higgsing, one of half D5-branes remains intact, while the other half D5-brane is recombined onto the two half D7-branes on the $\mathrm{O}5^-$-plane. So there are effectively three half D5-branes in between two half D7-branes so that the two half D5-branes can be Higgsed away.

\begin{figure}
\centering
\includegraphics[scale=1]{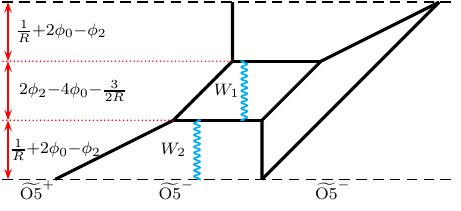}
\caption{A 5-brane web for 6d $SU(3)$ theory with $\mathbb{Z}_2$ twist. The half monodromy cut extends from the right to the left.} \label{fig:SU3Z2}
\end{figure}

Now, by taking each half D7-brane in the opposite directions, leaving the half D7 monodromy cut being presented from the right to the left, one obtains a 5-brane configuration given in Figure \ref{fig:SU3Z2}. 
The masses of W-bosons after the Higgsing are then given by
\begin{align}
m_{W_1} = 2\phi_2 - 4\phi_0 - \frac{3}{2R}\ , \qquad
m_{W_2} = \frac{1}{R} + 2\phi_0 - \phi_2\ , 
\end{align}
which form the affine Cartan matrix of twisted affine Lie algebra $A_2^{(2)}$.

\subsection{Partition function from 5-brane webs}
\begin{figure}
\centering
\begin{subfigure}{0.45\textwidth}
\centering
\includegraphics[scale=1]{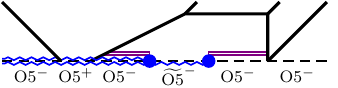}
\caption{}
\end{subfigure}
\begin{subfigure}{0.45\textwidth}
\centering
\includegraphics[scale=1]{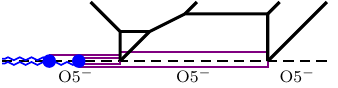}
\caption{}
\end{subfigure}
\caption{(a) Coupling a spinor $(-1, 1)$ 5-brane to $SU(3)$ theory with $\mathbb{Z}_2$ twist. (b) Applying the generalized flop transition and moving the half D7-branes and monodromy cuts to the right. Several half D5-branes are created due to the Hanany-Witten transition.} \label{fig:spinor_G2}
\end{figure}

With this 5-brane web, we are ready to compute the topological string partition function for the 6d $SU(3)$ gauge theory with $\mathbb{Z}_2$ twist. Like 5-brane configurations for the $SO(8)$ theory with  $\mathbb{Z}_2$ twist in the previous section, this 5-brane configuration is not smoothly connected to its reflected image over the orientifold plane. In order to make a desirable 5-brane configuration where the original 5-brane and its reflected mirror image are smoothly connected, we introduce an auxiliary spinor hypermultiplet on the 5-brane configuration, as done in the previous section. It allows us then to apply topological vertex formulation with an $O5$-plane to compute the partition function for the 6d $SU(3)$ gauge theory with $\mathbb{Z}_2$ twist, followed by taking the decoupling limit of the newly introduced hypermultiplet. For instance, see Figure~\ref{fig:spinor_G2}. The lower part of the 5-brane configuration for 6d $SU(3)$ gauge theory with $\mathbb{Z}_2$ twist is given in Figure~\ref{fig:spinor_G2}(a) where we introduced a hypermultiplet\footnote{From $G_2$ gauge theory point of view, this hypermultiplet is a fundamental hyper which originates from a hypermultiplet in the spinor representation of an $SO(7)$ gauge theory.} on the left hand side of the web configuration. Here two half D7-branes are still in the middle of the brane web which are denoted by blue dots and their monodromy cuts are the wiggly lines pointing to the left. For computational ease, we take these half D7-branes to the far left instead of taking them in the opposite directions. This makes a 5-brane configuration with only O5-plane, as depicted in Figure~\ref{fig:spinor_G2}(b). When taking all the half D7-branes to the left, half D5-branes are created as a result of the Hanany-Witten transitions, which is depicted in Figure~\ref{fig:spinor_G2}(b), where half D5-branes are solid lines in violet. In actual computation, they can be viewed as follows: one full external D5-brane (or two half D5-branes) is connected to the right $(0, 1)$ and $(1, 1)$ 5-branes, and an additional full D5-brane is connected to the left $(0, 1)$ and $(1, 1)$ 5-branes. 

\begin{figure}
\centering
\includegraphics[scale=1]{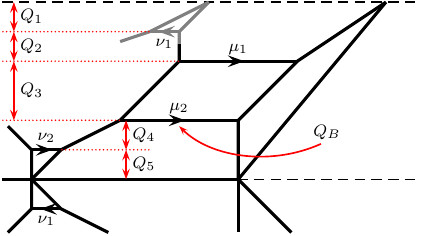}
\caption{A brane web of $SU(3)$ theory with $\mathbb{Z}_2$ twist coupled with two spinors. 
}
 \label{fig:6d_SU3Z2_vtx}
\end{figure}

\begin{figure}
\centering
\includegraphics[scale=1]{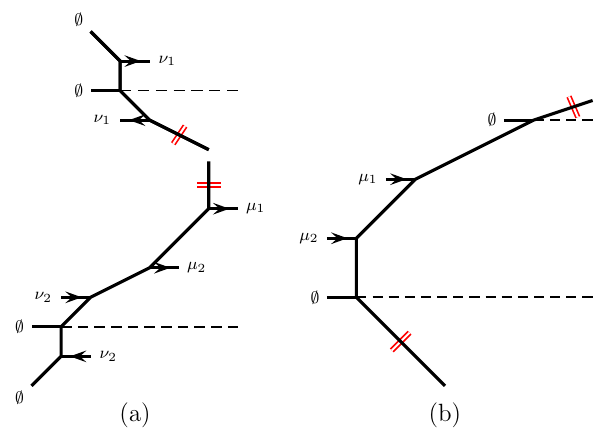}
\caption{Strip diagrams for 5-brane webs for $SU(3)$ theory with $\mathbb{Z}_2$ twist given in  Figure~\ref{fig:6d_SU3Z2_vtx}. The left part (a) and the right part (b) of the web diagram in Figure~\ref{fig:6d_SU3Z2_vtx}. $Z_{\mathrm{left}}$ is based on (a) and $Z_{\mathrm{right}}$ is based on (b). The edges denoted with ${\color{red}||}$ are identified.}\label{fig:6d_SU3Z2_LR}
\end{figure}

For the upper part of the 5-brane web in Figure \ref{fig:SU3Z2}, the 5-brane configuration is not smoothly connected to its mirror image as well. One may again couple another hypermultiplet as in Figure~\ref{fig:6d_SU3Z2_vtx}. It is also possible to use the symmetry of the 5-brane web. Instead of computing the partition function based on Figure~\ref{fig:6d_SU3Z2_vtx}, we use the following trick. We recycle the lower part of 5-brane configuration with a hyper for the upper part computation. In other words, when implementing the topological vertex for 5-brane configuration on the upper orientifold plane, we re-use the lower part configuration with a hyper as if it is a suitable 5-brane configuration for the upper part and perform the partition function computation by properly gluing 5-branes. For instance, as in Figure \ref{fig:6d_SU3Z2_LR}, we first consider strip diagrams. For the upper part, in particular, we use the strip diagram of the lower part and glue the upper and lower part together. In the figure, the two edges marked with ${\color{red}||}$ symbols are identified. This trick is not harmful, as the point of introducing extra hypermultiplets is to arrange the 5-brane configuration in a way that we can apply the topological vertex. Moreover, we decouple the contribution of these newly introduced hypermultiplets.

It follows then that the the strip diagram on  Figure~\ref{fig:6d_SU3Z2_LR}(a) yields
\begin{align}
&Z_{\mathrm{left}}(\mu_1, \mu_2, \nu_1, \nu_2)  \\
&= g^{\frac{1}{2}(\norm*{\mu_1}^2 + \norm*{\mu_2}^2 + 2\norm*{\nu_1}^2 + 2\norm*{\nu_2^t}^2)} \tilde{Z}_{\mu_1}(g) \tilde{Z}_{\mu_2}(g) \tilde{Z}_{\nu_1}(g)^2 \tilde{Z}_{\nu_2}(g)^2 \frac{\mathcal{R}_{\emptyset \mu_1}(Q_{3,4,5}) \mathcal{R}_{\emptyset \mu_1^t}(Q_{1,2})}{\mathcal{R}_{\emptyset \emptyset}(Q_{1,2,3,4,5})} \nonumber \\
& \quad \times \frac{  \mathcal{R}_{\emptyset \mu_2}(Q_{4,5}) \mathcal{R}_{\emptyset \mu_2^t}(Q_{1,2,3})  \mathcal{R}_{\nu_1 \emptyset}(Q_{1,1,2,3,4,5}) \mathcal{R}_{\emptyset \nu_2}(Q_{1,2,3,4,5,5}) \mathcal{R}_{\mu_1^t \nu_1^t}(Q_2) \mathcal{R}_{\mu_2^t \nu_1^t}(Q_{2,3})}{ \mathcal{R}_{\emptyset \nu_1^t}(Q_{2,3,4,5}) \mathcal{R}_{\emptyset \nu_2^t}(Q_{1,2,3,4}) \mathcal{R}_{\mu_1 \mu_2^t}(Q_3) \mathcal{R}_{\mu_1^t \nu_1}(Q_{1,1,2}) } \nonumber \\
& \quad\times \frac{ \mathcal{R}_{\mu_1 \nu_2^t}(Q_{3,4}) \mathcal{R}_{\mu_2 \nu_2^t}(Q_4) \mathcal{R}_{\nu_1 \nu_1}(Q_{1,1}) \mathcal{R}_{\nu_1^t \nu_2}(Q_{2,3,4,5,5}) \mathcal{R}_{\nu_2 \nu_2}(Q_{5,5}) \mathcal{R}_{\nu_1 \nu_2^t}(Q_{1,1,2,3,4}) }{ \mathcal{R}_{\mu_2^t \nu_1 }(Q_{1,1,2,3}) \mathcal{R}_{\mu_1 \nu_2}(Q_{3,4,5,5}) \mathcal{R}_{\mu_2 \nu_2}(Q_{4,5,5}) \mathcal{R}_{\nu_1 \nu_2}(Q_{1,1,2,3,4,5,5}) \mathcal{R}_{\nu_1^t \nu_2^t}(Q_{2,3,4})},\nonumber
\end{align}
where we used the K\"ahler parameters $Q_i$ assigned in Figure \ref{fig:6d_SU3Z2_vtx} with the convention that $Q_{i,j,\cdots} = Q_{i} Q_j \cdots$.

The strip on Figure~\ref{fig:6d_SU3Z2_LR}(b) can be treated in a similar way except that the upper and lower edges marked with ${\color{red}||}$ symbol are identified.

The topological string partition function for the strip in Figure~\ref{fig:6d_SU3Z2_LR}(b) is a bit involving but can be obtained in a straightforward manner following \cite{Haghighat:2013gba, macdonald_symmetric_1995}. One needs to repeatedly use various Cauchy identities in Appendix \ref{sec:special}. The contribution of the strip diagram in Figure~\ref{fig:6d_SU3Z2_LR}(b) then reads \begin{align}\label{eq:SU3Z2_right}
&Z_{\mathrm{right}}(\mu_1, \mu_2) 
= g^{\frac{1}{2}(\norm*{\mu_1^t}^2 + \norm*{\mu_2^t}^2)} \tilde{Z}_{\mu_1}(g) \tilde{Z}_{\mu_2}(g) \times\nonumber \\
& \quad \times \prod_{m=1}^{\infty} \frac{1}{1-q^m} \frac{1}{\mathcal{R}_{\emptyset \emptyset }(q^{2 m})^2 \mathcal{R}_{\emptyset \emptyset }(q^{2 m-1/2})^2 \mathcal{R}_{\emptyset \emptyset}(q^{2 m-1})^2 \mathcal{R}_{\emptyset \emptyset }(q^{2 m-3/2})^2  } \nonumber \\
& \quad \times \frac{1}{\mathcal{R}_{\emptyset \mu_1}(q^{2 m-2} Q_{3,4,5}) \mathcal{R}_{\emptyset \mu_1}(q^{2 m-1} Q_{3,4,5}) \mathcal{R}_{\emptyset \mu_1}(q^{2 m} Q_{1,2}^{-1}) \mathcal{R}_{\emptyset \mu_1}(q^{2 m-1} Q_{1,2}^{-1})} \nonumber \\
& \quad\times  \frac{1}{ \mathcal{R}_{\emptyset \mu_1^t}(q^{2 m-2} Q_{1,2}) \mathcal{R}_{\emptyset \mu_1^t}(q^{2 m-1} Q_{1,2}) \mathcal{R}_{\emptyset \mu_1^t}(q^{2 m-2} Q_{1,1,2,2,3,4,5}) \mathcal{R}_{\emptyset \mu_1^t }(q^{2m-1} Q_{1,1,2,2,3,4,5})} \nonumber \\
& \quad \times \frac{1}{\mathcal{R}_{\emptyset \mu_2}(q^{2m-2} Q_{4,5}) \mathcal{R}_{\emptyset \mu_2}(q^{2 m-1} Q_{4,5}) \mathcal{R}_{\emptyset \mu_2}(q^{2 m} Q_{1,2,3}^{-1}) \mathcal{R}_{\emptyset \mu_2 }(q^{2m-1} Q_{1,2,3}^{-1})} \nonumber \\
& \quad\times  \frac{1}{ \mathcal{R}_{\emptyset \mu_2^t }(q^{2 m} Q_{4,5}^{-1}) \mathcal{R}_{\emptyset \mu_2^t}(q^{2 m-1} Q_{1,2,3}) \mathcal{R}_{\emptyset \mu_2^t }(q^{2 m-1} Q_{4,5}^{-1}) \mathcal{R}_{\emptyset \mu_2^t }(q^{2 m-2} Q_{1,2,3})} \nonumber \\
& \quad\times  \frac{1}{\mathcal{R}_{\mu_1 \mu_1^t}(q^{2 m}) \mathcal{R}_{\mu_1 \mu_1^t}(q^{2 m-1})  \mathcal{R}_{\mu_2 \mu_2^t}(q^{2m}) \mathcal{R}_{\mu_2 \mu_2^t}(q^{2 m-1})} \nonumber \\
& \quad\times  \frac{1}{\mathcal{R}_{\mu_1^t \mu_2}(q^{2 m} Q_3^{-1}) \mathcal{R}_{\mu_1^t \mu_2}(q^{2 m-1} Q_3^{-1}) \mathcal{R}_{\mu_1 \mu_2^t}(q^{2 m-1} Q_3)  \mathcal{R}_{\mu_1 \mu_2^t}(q^{2 m-2} Q_3) }\ ,
\end{align}
where $q=(Q_1 Q_2 Q_3 Q_4 Q_5)^2$.  

The full topological string partition function $Z$ is obtained by combining $Z_{\mathrm{left}}$ and $Z_{\mathrm{right}}$, 
\begin{align}\label{eq:SU3Z2_Z}
Z = \sum_{\mu_i, \nu_i} (-Q_B)^{\abs{\mu_1}+\abs{\mu_2}} Q_1^{\abs{\nu_1}} Q_5^{\abs{\nu_2}} f_{\mu_1} f_{\mu_2}^{-1} f_{\nu_1}^3  f_{\nu_2} Z_{\mathrm{left}}(\mu_1, \mu_2, \nu_1, \nu_2) Z_{\mathrm{right}}(\mu_1, \mu_2).
\end{align}
It is convenient to further rescale the scalars in Figure~\ref{fig:SU3Z2} as
\begin{align}\label{eq:SU3Z2_reparametrization}
\phi_2 \to \phi_2 + 2\phi_0 + \frac{3}{4R}\ , 
\end{align}
which gives that the K\"ahler parameters are expressed in terms of physical parameters
\begin{align}
- \log Q_1 Q_2 &= -\log Q_4 Q_5 = -\phi_2 + \frac{\tau}{4}\ , \nonumber\\
- \log Q_3 &= 2\phi_2\ , \cr
- \log Q_B &=  \log u + 2\phi_2\ , 
\end{align}
where 
$u$ is the string fugacity. We note that the relation $q=(Q_1 Q_2 Q_3 Q_4 Q_5)^2 = e^{-\tau}$ is not changed under the rescaling. With  $A = e^{- \phi_2}$, we can write the K\"ahler parameters $Q_2$ and $Q_4$ as 
$Q_2 = q^{1/4} A^{-1} Q_1^{-1}$ and $Q_4 = q^{1/4} A^{-1} Q_5^{-1}$ where $Q_1, Q_5$ are the K\"ahler parameters associated with the spinors which we will decouple.

The perturbative partition function $Z_{\mathrm{pert}}$ is obtained by setting $\mu_1 = \mu_2 = \emptyset$ in  \eqref{eq:SU3Z2_Z} and by summing up over the Young diagrams $\nu_i$, 
as a function of $g$, $A$, $q$, $Q_1$ and $Q_5$. As an expansion of $q$, 
we find the perturbative part is given by
\begin{align}\label{eq:SU3Z2_pert}
Z_{\mathrm{pert}}\!
= \operatorname{PE} \!\bigg[\frac{2g A^2}{(1\!-g)^2} \!+ \frac{g(2A^{-1}\! + 2A) q^{1/4}}{(1-g)^2}\! + \frac{4g q^{1/2}}{(1\!-g)^2} \!+ \frac{2g(A \!+  A^{-1}) q^{3/4}}{(1-g)^2}\! +  \cdots \!\bigg],
\end{align}
where the $\cdots$ denotes the terms involving $Q_1$ and $Q_5$ which we will decouple later.  
This is expected result as in Appendix~\ref{sec:affine}. If we take the large radius limit corresponding to $q\to 0$, the states depending on KK-momentum will be truncated, and as a result, only $q^0$ term will remain.

The partition function for 6d self-dual strings $Z_{\mathrm{string}}$ is obtained by summing up over all Young diagrams in \eqref{eq:SU3Z2_Z}, expand about $q$ and $A$, and also by decoupling the auxiliary spinor hypermultiplets that we introduced for computational ease, $Q_1, Q_5 \to \infty$ as discussed in the previous section,  
\begin{align}
Z_{\mathrm{string}} = \frac{Z}{Z_{\mathrm{pert}}} = 1 + u Z_1 + u^2 Z_2 + \cdots,
\end{align}
where $Z_n$ correspond to the $n$-string elliptic genus, 
\begin{align}\label{eq:Z1Z2 for SU3_9 from 5-brane}
Z_1 &= \frac{2g A^2}{(1-g)^2 (1-A^2)^2} + \frac{2gA(1+A^2)}{(1-g)^2 (1-A^2)^2} q^{1/4} + \frac{6g A^2}{(1-g)^2 (1-A^2)^2} q^{1/2} \nonumber \\
& \quad + \frac{6g A(1+A^2)}{(1-g)^2 (1-A^2)^2} q^{3/4} + \frac{6 A^4 g + 2A^2 (2g^2 + 7g + 2) + 6g}{(1-g)^2 (1-A^2)^2} q + O(q^{5/4})\ , \nonumber\\
Z_2 &= \frac{g^4 A^4 \Big(A^4 (3g^2+2g+3) - 2A^2 (g^2+6g+1) + 3g^2 + 2g + 3\Big)}{(1-g)^4 (1+g)^2 (1-A^2)^2 (A^2-g)^2 (1-A^2 g)^2} \nonumber \\
&\quad + \frac{2 g^3 A^3 (A^2+1) \big(2 A^4 g + A^2 (g^2-6g+1) + 2g\big)}{(1-g)^4 (1-A^2)^2 (A^2-g)^2 (1-A^2g)^2} q^{1/4} \nonumber \\
&\quad + \frac{A^2 g^3 \Big(A^4 (3g^2+2g+3) - 2A^2 (g^2+6g+1) + 3g^2 + 2g + 3\Big)
}{(1-g)^4 (1+g)^2 (1-A^2)^2 (A^2-g)^2 (1-A^2 g)^2} \times \nonumber \\
&\qquad \times \big(A^4 g + 2A^2(g+1)^2 + g\big) q^{1/2} + O(q^{3/4})\ . 
\end{align}
Though we presented the result of one- and two-string elliptic genus only, higher $n$-string elliptic genus can be computed in a straightforward manner.

\subsection{Elliptic genus by Higgsing 6d \texorpdfstring{$G_2+1\mathbf{F}$}{G2+1F}}\label{sec:G2+1Feg}

In this subsection, we compute the elliptic genus  to cross-check the partition for the $SU(3)$ theory with $\mathbb{Z}_2$ twist obtained from 5-brane webs in the previous subsection. 
As discussed, the $SU(3)$ theory with $\mathbb{Z}_2$  twist can be obtained by Higgsing the 6d $G_2$ gauge theory with one fundamental hypermultiplet ($G_2+1\mathbf{F}$). We hence start by computing the elliptic genus for the 6d $G_2$ gauge theory with one fundamental hypermultiplet and apply the Higgsing that leads to the $SU(3)$ theory with $\mathbb{Z}_2$ twist.

The perturbative part of the partition function comes from the contributions from the vector multiplet and the hypermultiplet,  
$Z_{\mathrm{pert}}^{G_2+1\mathbf{F}}=Z_{\mathrm{pert}}^{\mathrm{gauge}}Z_{\mathrm{pert}}^{\mathrm{hyper}}$. As we have the correspond 5-brane web in Figure~\ref{fig:G2_flop}, we can readily obtain the perturbative part.  
The vector multiplet contribution to the perturbative part for $G_2+1\mathbf{F}$ takes the following form
\begin{align}
Z_{\mathrm{pert}}^{\mathrm{gauge}}
= \operatorname{PE}\qty[ \frac{2g}{(1-g)^2} \qty( \chi_{\Delta_+}^{G_2} + q \chi_{\Delta_-}^{G_2}) \frac{1}{1-q} ]\ ,
\end{align}
where 
$\chi_{\Delta_\pm}^{G_2}$ are the positive and negative parts from the characters for the adjoint representation of $G_2$. The hypermultiplet contribution to the perturbative part is given by
\begin{align}
Z_{\mathrm{pert}}^{\mathrm{hyper}}\!
=\! \operatorname{PE}\!\qty[ -\frac{g}{(1-g)^2} \qty(M \!+\! \frac{q}{M}) \qty(\frac{x_1}{q}\! + \frac{x_1}{x_2}\! + \frac{x_2}{x_1^2}\! + 1\! + \frac{x_1^2}{x_2} \!+ \frac{x_2}{x_1}\! + \frac{q}{x_1}) \frac{1}{1-q} ] ,
\end{align}
where 
$M = e^{- m}$ and $x_i = e^{-\phi_i}$. 
These perturbative part can be easily computed from the web diagram in Figure~\ref{fig:G2_flop}. 

To perform the Higgsing leading to the $SU(3)$ gauge theory with $\mathbb{Z}_2$ twist, we give a vev to the hypermultiplet carrying KK momentum. For that, we recover the affine node $\phi_0$ as in Figure~\ref{fig:G2_flop}, and impose the Higgsing conditions \eqref{eq:G2_higgsing}. By rescaling the scalars \eqref{eq:SU3Z2_reparametrization}, we obtain the perturbative part of the partition function for the $SU(3)$ gauge theory with $\mathbb{Z}_2$ twist:
\begin{align}
Z_{\mathrm{pert}}\!=\!\operatorname{PE}\!\bigg[ \frac{2g}{(1-g)^2} & \Big(\! (A^2\! -\! 2) + (A\!+\! A^{-1}) q^{1/4}  + q^{1/2} + (A\! +\! A^{-1}) q^{3/4} + \!\cdots\! \Big) \bigg],
\end{align}
where
$A = x_2 = e^{-\phi_2}$. This result is the same as that obtained from the topological vertex calculation \eqref{eq:SU3Z2_pert} up to the Cartan parts. 

\begin{figure}
\centering
\includegraphics[scale=1]{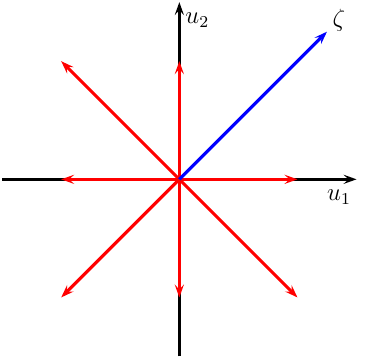}
\caption{An auxiliary vector  $\zeta$ for the JK-residue calculation of two-string elliptic genus of the $G_2$ gauge theory with one fundamental hypermultiplet.} \label{fig:G2_JK}
\end{figure}

We now compute the elliptic genus of the 6d $G_2$ theory with one fundamental hypermultiplet, following~\cite{Kim:2018gjo}. 
The symmetries of the 2d worldsheet theory on self-dual strings are $U(n)$ gauge symmetry for string number $n$, $SU(3)$ global symmetry which is enhanced to $G_2$ in IR, $U(1)_{\sf J}$ and $SU(2)_l$ global symmetries. Here, for $SU(2)_R$ R-symmetry and $SO(4) = SU(2)_l \times SU(2)_r$ rotation symmetry of the transverse $\mathbb{R}^4$, the charge ${\sf J}$ is identified as the sum of the $SU(2)_r$ charge and the $SU(2)_R$ charge,  ${\sf J} = J_r + J_R$.
The worldsheet fields can be written in $\mathcal{N}=(0, 2)$ multiplets.
There are the vector multiplet $V$, the Fermi multiplets $\lambda$, $(\hat{\lambda}, \check{\lambda})$, $(\Psi, \bar{\Psi})$, and the chiral multiplets $(q, \tilde{q})$, $(a, \tilde{a})$, $(\phi_i, \phi_4)$, $(b, \tilde{b})$. Their charges under the symmetries in the 2d worldsheet theory are summarized in Table~\ref{table:G2}.
\begin{table}
\centering
\begin{tabular}{|c|c|c|c|c|}
\hline
 & $U(n)$ & $SU(3)$ & $SU(2)_l$ & $U(1)_{\sf J}$ \\ \hline
$V$ & $\mathbf{adj}$ & $\mathbf{1}$ & $\mathbf{1}$ & $0$ \\ \hline
$\lambda$ & $\mathbf{adj}$ & $\mathbf{1}$ & $\mathbf{1}$ & $-1$ \\ \hline
$(q, \tilde{q})$ & $(\mathbf{n}, \bar{\mathbf{n}})$ & $(\bar{\mathbf{3}}, \mathbf{3})$ & $\mathbf{1}$ & $1/2$ \\ \hline
$(a, \tilde{a})$ & $\mathbf{adj}$ & $\mathbf{1}$ & $\mathbf{2}$ & $1/2$ \\ \hline
$(\phi_i, \phi_4)$ & $\bar{\mathbf{n}}$ & $(\bar{\mathbf{3}}, \mathbf{1})$ & $\mathbf{1}$ & 1/2 \\ \hline
$(b, \tilde{b})$ & $\overline{\mathbf{anti}}$ & $\mathbf{1}$ & $\mathbf{2}$ & $1/2$ \\ \hline
$(\hat{\lambda}, \check{\lambda})$ & $\overline{\mathbf{sym}}$ & $\mathbf{1}$ & $\mathbf{1}$ & $(0, -1)$ \\ \hline
$(\Psi, \tilde{\Psi})$ & $(\mathbf{n}, \bar{\mathbf{n}})$ & $\mathbf{1}$ & $\mathbf{1}$ & $0$ \\ \hline
\end{tabular}
\caption{The $\mathcal{N}=(0, 2)$ multiplets and symmetries in ADHM formalism of 6d $G_2$ gauge theory.} \label{table:G2}
\end{table}
Using the matter content in Table \ref{table:G2}, one can write down the $n$-string elliptic genus: 
\begin{align}
Z_n &= \frac{1}{n!} \frac{1}{(2\pi i)^n} \oint \prod_{I=1}^n du_I \cdot \Big(\frac{2\pi \eta^2}{i}\Big)^n \qty(\prod_{I \neq J}^n \frac{i \theta_1(u_I - u_J)}{\eta}) \qty(\prod_{I, J=1}^n \frac{i \theta_1(-2\epsilon_+ + u_I - u_J)}{\eta}) \nonumber \\
&\quad \times\qty(\prod_{I=1}^n \prod_{J=1}^{3} \frac{i^2 \eta^2}{\theta_1(\epsilon_+ \pm (u_I - v_J))}) \qty(\prod_{I, J=1}^n \frac{i^2 \eta^2}{\theta_1(\epsilon_{1,2} + u_I - u_J)}) \nonumber \\
&\quad \times\qty(\prod_{I=1}^n \prod_{J=1}^3 \frac{i \eta}{\theta_1(\epsilon_+ - u_I - v_J)} \frac{i \eta}{\theta_1(\epsilon_+ - u_I)}) \qty(\prod_{I < J}^n \frac{i \eta}{\theta_1(\epsilon_{1,2} - u_I - u_J)}) \nonumber \\
&\quad \times\qty(\prod_{I \leq J}^n \frac{i \theta_1(u_I + u_J)}{\eta} \frac{i \theta_1(-2\epsilon_+ + u_I + u_J)}{\eta}) \qty(\prod_{I=1}^n \frac{i^2 \theta_1(\pm u_I + m)}{\eta^2}).
\end{align}
Here, $u_I$ are the $U(n)$ parameters, $\eta$ is the Dedekind eta function, $v_I$ are the $SU(3)$ parameters subject to $v_1 + v_2 + v_3 = 0$. 
First consider the $n=1$ case. We choose an auxiliary vector $\zeta = (1)$. Then contributing poles are from $\epsilon_+ + u - v_J = 0$. The contour integral converts to the JK-residue sum. We thus obtain the one-string elliptic genus for the $G_2$ gauge theory with one hypermultiplet
\begin{align}\label{eq:G2_1string_genus}
Z_1^{G_2+1\mathbf{F}}  \! =\! \sum_{I=1}^3 \frac{\eta^6\, \theta_1(2v_I - 4\epsilon_+)\, \theta_1\!(m \pm (\epsilon_+ \!-\! v_I))}{\theta_1(\epsilon_{1,2}) \,\theta_1(2\epsilon_+ \!-\! v_I) \displaystyle{\prod_{J\neq I}} \theta_1(v_I\! -\! v_J) \,\theta_1(2\epsilon_+ \!-\! v_I \!+ v_J)\, \theta_1(2\epsilon_+ \!+\! v_J)},
%
\end{align}
where the $SU(3)$ condition $v_1 + v_2 + v_3 = 0$ is used. For the $n=2$ case, we choose an auxiliary vector $\zeta$ as in Figure~\ref{fig:G2_JK}.
The poles which survive in the JK-residue sum are
\begin{align}\label{eq:G2_poles}
(i) \left\{
\begin{array}{l}
\epsilon_+ + u_2 - v_I = 0 \\
\epsilon_+ + u_1 - v_J = 0
\end{array}\right.
&&
(ii) \left\{
\begin{array}{l}
\epsilon_+ + u_2 - v_J = 0 \\
\epsilon_{1,2} + u_1 - u_2 = 0
\end{array}\right.
&&
(iii) \left\{
\begin{array}{l}
\epsilon_{1,2} + u_2 - u_1 = 0 \\
\epsilon_+ + u_1 - v_J = 0\ .
\end{array}\right.
\end{align}
The first poles $(i)$ in \eqref{eq:G2_poles} give
\begin{align}
\mathrm{Res}_1
&=  \sum_{I\neq J}^3 \frac{\eta^{12} \theta_1(2v_{I,J}-4\epsilon_+) \theta_1(\epsilon_+-v_{I,J}\pm m)}{2\,  \theta_1\!(\epsilon_{1,2})^2\,  \theta_1\!(2\epsilon_+ \pm v_{I,J}) \,\theta_1\!(\epsilon_{1,2} \pm (v_I - v_J)
)}  \\
& \quad\! \times\!\!\prod_{K \neq I, J}\! \frac{\theta_1(4\epsilon_+ + v_K)}{\theta_1\!(2\epsilon_+ + v_K)\, \theta_1(3\epsilon_+ \pm \epsilon_-\!+\! v_K) \,\theta_1\!(v_{I,J}\! - v_K)\,  \theta_1\!(2\epsilon_+\! - v_{I,J}\! + v_K)} .\nonumber
\end{align}
The second poles $(ii)$ give
\begin{align}
& \mathrm{Res}_2 \nonumber \\
&= \!\frac{\eta^{12}}{2} \sum_{I=1}^3 \!\Bigg( \!\frac{\theta_1(5\epsilon_+\! + \!\epsilon_- \!-2v_I) \theta_1(6\epsilon_+ \!+\! 2\epsilon_- \!-\! 2v_I) \theta_1(\epsilon_+ \!-\! v_I \pm\! m) \theta_1(2\epsilon_+\! +\! \epsilon_-\! -\! v_I \!\pm m)}{\theta_1(\epsilon_{1,2}) \theta_1(2\epsilon_1) \theta_1(-2\epsilon_-) \theta_1(2\epsilon_+ - v_I) \theta_1(3\epsilon_+ + \epsilon_- - v_I)} \nonumber \\
& \quad  \qquad \times\prod_{J \neq I} \frac{1}{\theta_1(v_I - v_J) \theta_1(\epsilon_1 - v_I + v_J) \theta_1(2\epsilon_+ - v_I + v_J) \theta_1(3\epsilon_+ + \epsilon_- - v_I + v_J)} \nonumber \\
&\quad \qquad \quad \times\frac{1}{\theta_1(2\epsilon_+ + v_J) \theta_1(3\epsilon_+ + \epsilon_- + v_J)} + (\epsilon_1 \leftrightarrow \epsilon_2, \: \epsilon_- \to -\epsilon_-) \bigg)\ .
\end{align}
Notice that the third poles (\textit{iii}) can be obtained by $u_1 \leftrightarrow u_2$ in $(ii)$. It follows that the residue for (\textit{iii}) is the same as that for $(ii)$, $\mathrm{Res}_3=\mathrm{Res}_2$. The two-string elliptic genus of the $G_2$ theory with one hypermultiplet is then given by
\begin{align}\label{eq:G2_2string_genus}
Z_2^{G_2+1\mathbf{F}} = \mathrm{Res}_1 + 2 \times \mathrm{Res}_2\ .
\end{align}

We now Higgs the 6d $G_2$ theory. The procedure is similar to that for the $SO(10)$ case in the previous section. As the elliptic genus results are written in terms of the $SU(3)$ parameters, rather than the $G_2$ parameters, some additional implementation is required for the proper Higgs from the 6d $G_2$ theory to the $SU(3)$ theory with $\mathbb{Z}_2$ twist.

We begin by identifying the relation between the $SU(3)$ parameters $v_i$ and the $G_2$ fundamental weights. 
It follows from the embedding  
\begin{align}
	G_2 &\supset SU(3)\crcr
	{\bf 7} &= {\bf 1+3+\bar 3}\, 
\end{align}
that the character for the fundamental representation of $G_2$ is expressed in terms of the $SU(3)$ characters, 
$1 + \sum_I (e^{ v_i}+e^{- v_I})$, and the parameter map between $SU(3)$ parameters $v_i$ and the $G_2$ fundamental weights is $v_1 \to \phi_1$, $v_2 \to \phi_1 - \phi_2$ and $v_3 \to \phi_2 - 2\phi_1$.  
By applying the Higgsing conditions from the 5-brane web diagram \eqref{eq:G2_higgsing} as well as the reparameterization \eqref{eq:SU3Z2_reparametrization}, one finds that the proper Higgsing from the $G_2$ theory to the $SU(3)$ gauge theory with $\mathbb{Z}_2$ twist is given by 
\begin{align}
v_1 \to \frac{\tau}{2}, \qquad v_2 \to -\phi_2 - \frac{\tau}{4}, \qquad v_3 \to \phi_2 - \frac{\tau}{4}, \qquad m \to \frac{\tau}{2}.
\end{align}
Substituting these into the one-string elliptic genus for the $G_2$ theory given in \eqref{eq:G2_1string_genus}, we get the one-string elliptic genus for the $SU(3)$ gauge theory with $\mathbb{Z}_2$ twist:
\begin{align}\label{eq:SU3Z2_Z1}
Z_1 = \frac{2 \, \eta^6\,  \theta_1(-2\phi_2 + \frac{\tau}{2})}{\theta_1(\epsilon_-)^2 \theta_1(2\phi_2)^2 \theta_1(\frac{\tau}{2}) \theta_1(-\phi_2 + \frac{\tau}{4}) \theta_1(\phi_2 - \frac{3 \tau}{4})}\ .
\end{align}
It follows that two-string elliptic genus for the $SU(3)$ gauge theory with $\mathbb{Z}_2$ twist is given by
\begin{align}\label{eq:SU3Z2_Z2}
Z_2 &= \frac{\eta^{12}\, \theta_1(\pm 2\phi_2 + \frac{\tau}{2})}{\theta_1(\epsilon_-)^4 \theta_1(\epsilon_- \pm 2 \phi_2)^2 \theta_1(\pm \phi_2 + \frac{\tau}{4}) \theta_1(\pm \epsilon_- + \frac{\tau}{2}) \theta_1(\pm \phi_2 + \frac{3 \tau}{4})} \nonumber \\
& \quad  + \bigg( \frac{2 \eta^{12} \theta_1(\epsilon_- - 2\phi_2 + \frac{\tau}{2}) \theta_1(2\epsilon_- - 2\phi_2 + \frac{\tau}{2})}{\theta_1(\epsilon_-)^2 \theta_1(2\epsilon_-)^2 \theta_1(2\phi_2)^2 \theta_1(\epsilon_- - 2\phi_2)^2 \theta_1(-\phi_2 + \frac{\tau}{4}) \theta_1(\epsilon_- - \phi_2 + \frac{\tau}{4})} \nonumber \\
& \qquad \quad \times \frac{1}{\theta_1(\frac{\tau}{2}) \theta_1(\epsilon_- + \frac{\tau}{2}) \theta_1(-\phi_2 + \frac{3\tau}{4}) \theta_1(\epsilon_- - \phi_2 + \frac{3\tau}{4})} + (\epsilon_- \to -\epsilon_-) \bigg).
\end{align}
It is straightforward to see that these one- and two-string elliptic genera, \eqref{eq:SU3Z2_Z1} and \eqref{eq:SU3Z2_Z2}, agree with the $Z_1$ and $Z_2$ obtained from topological vertex given in \eqref{eq:Z1Z2 for SU3_9 from 5-brane}, by double expanding \eqref{eq:SU3Z2_Z1} and \eqref{eq:SU3Z2_Z2} in terms of $q$ and $A$ and also taking the unrefined limit.

\section{Conclusion} \label{sec:conclusion}
In this paper, we computed the partition functions of 6d $SO(8)$ and $SU(3)$ gauge theories with $\mathbb{Z}_2$ outer automorphism twist in two different perspectives. One is to use their Type IIB 5-brane webs where 6d conformal matter theories of D-type gauge symmetry can be engineered as 5-brane webs with two $\mathrm{O}5$-planes. Among various RG flows, we discussed the Higgsing procedure on 5-brane webs giving rise to the 6d $SO(8)$ and $SU(3)$ gauge theories with $\mathbb{Z}_2$ twist, from the 6d  $SO(10)$ gauge theory with two flavors and from the 6d  $G_2$ gauge theory with a flavor, respectively. We computed the partition functions of these theories following the topological vertex formalism in the presence of $\mathrm{O}5$-planes developed in \cite{Kim:2017jqn}, by introducing and decoupling spinor matter fields to implement the topological vertex as done \cite{Hayashi:2018bkd}. The other is to directly apply the outer automorphism twists on the elliptic genera for the $SO(10)$ and $G_2$ gauge theories, by Higgsing the hypermultiplets with proper KK-momentum dependence. We checked that the partition functions based on topological vertex with O5-planes agree with the elliptic genera after Higgsings. As the elliptic genera are fully refined, we compared them in the unrefined limit as well as by double expanding each in terms of the KK momentum and the Coulomb branch parameters.

When computing the elliptic genus for the $SU(3)$ theory with $\mathbb{Z}_2$ twist in section \ref{sec:G2+1Feg}, one may wonder whether one can apply the twisting directly on the elliptic genus for the $SU(3)$ theory, implementing the KK-momentum shifts as done for the $SO(8)$ case in section \ref{sec:directTwistingSO8}. In the case of $SO(8)$ theory with $\mathbb{Z}_2$ twist, the order two outer automorphism maps fundamental representation to itself, and hence KK-momentum shifts can be easily understood. For $\mathfrak{su}(3)$ algebra, however, the fundamental representation $\mathbf{3}$ maps to the anti-fundamental representation $\bar{\mathbf{3}}$ under the order two outer automorphism. From the ADHM perspective, there are one chiral multiplet which transforms as $\mathbf{3}$ and two chiral multiplets which transform as $\bar{\mathbf{3}}$, and hence it is not clear how to perform the automorphism twist on these $SU(3)$ states in the ADHM construction. It would be good if a more systematic study along this direction is carried out so that it would be even applicable to order three outer automorphism twist.

Yet as another independent check, we computed, in~\cite{Kim:2020hhh}, the BPS spectrum of the 6d theories with $\mathbb{Z}_2$ twist using the bootstrap method via the Nakajima-Yoshioka's blowup equations, which provides fully refined partition functions. We checked our results against the BPS spectrum from the blowup formula and found that two results completely match.

\section*{Acknowledgements}
We would like to thank Kimyeong Lee, Kaiwen Sun, Xing-Yue Wei, and  Futoshi Yagi for useful discussions. 
S.K. thanks APCTP, KIAS, and POSTECH for hospitality for his visit. The research of HK and MK is supported by the POSCO Science Fellowship of POSCO TJ Park Foundation and the National Research Foundation of Korea (NRF) Grant 2018R1D1A1B07042934.

\appendix
\section{Twisted boundary condition and affine Lie algebras} \label{sec:affine}

\begin{figure}
\centering
\includegraphics[scale=1]{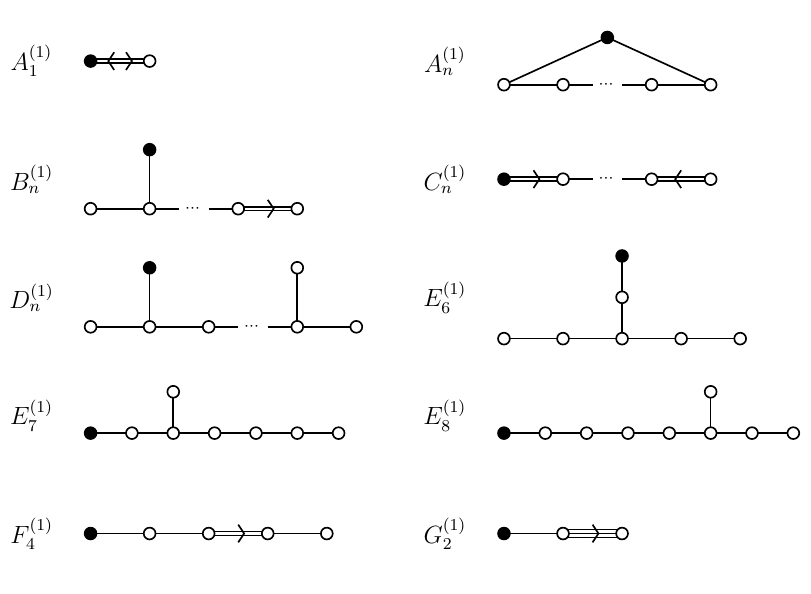}
\caption{Dynkin diagrams of untwisted affine Lie algebras, where the black node $\bullet$ represents the affine node.} \label{fig:dynkin-affine}
\end{figure}

Consider a 6d gauge theory with a simple Lie algebra $\mathfrak{g}$ and compactification of the theory on a circle $S^1$ of radius $R$. 
For the gauge fields $A_\mu = A_{\mu}^a T^a$, where $T^a$ lie in the adjoint representation of $\mathfrak{g}$, we impose the periodic boundary condition
\begin{align}\label{eq:periodic bc}
A_{\mu}(x^i, x^6 + 2\pi R) = A_\mu(x^i, x^6)\ .
\end{align}
Here, $i=1, 2, \cdots, 5$ and $x^6$ is the coordinate along $S^1$.  
Fourier expansion of the gauge fields preserving the periodic boundary condition takes the form 
\begin{align}
A_\mu(x^i, x^6) = A_{\mu}^a(x^i, x^6) T^a = \sum_{n\in \mathbb{Z}} e^{i \frac{n x^6}{R}} A_{\mu, n}^a(x^i) T^a = \sum_{n\in \mathbb{Z}} A_{\mu, n}^a(x^i) T^a_n,
\end{align}
where $T^a_n = T^a e^{i n x^6 / R}$. The new basis $T^a_n$ satisfies
\begin{align}
\comm{T^a_n}{T^b_m} = f^{abc}\, T^c_{n+m}\ ,
\end{align}
where $f^{abc}$ is the structure constant of $\mathfrak{g}$, $\comm*{T^a}{T^b} = f^{abc} T^c$. The basis $\{T^a_n\}$ with possible central extensions generates an untwisted affine Lie algebra. For a simple Lie algebra of type $X_\ell$, the untwisted affine Lie algebra constructed using the periodic boundary condition \eqref{eq:periodic bc} is denoted by $X_\ell^{(1)}$, and their Dynkin diagram is shown in Figure~\ref{fig:dynkin-affine}. The black circles are affine nodes. Excluding it gives the Dynkin diagram of corresponding simple Lie algebra $X_\ell$.

Instead of the periodic boundary condition, one can impose a twisted boundary condition
\begin{align}\label{eq:twistedbc}
A_\mu (x^i, x^6 + 2\pi R) = \sigma(A_\mu (x^i, x^6))\ ,
\end{align}
where $\sigma$ is a finite order automorphism of $\mathfrak{g}$. If $ \sigma $ is an order $ m $ automorphism, i.e., $\sigma^m = 1$, then the Fourier expansion of the gauge fields preserving the twisted boundary condition \eqref{eq:twistedbc} is
\begin{align}
A_\mu(x^i, x^6) = \sum_{n, k} e^{i \frac{x^6}{R}(n+\frac{k}{m})} A_{\mu, n, k}^a(x^i) T^a = \sum_{n, k} A_{\mu, n, k}^a T^a_{n+k/l}\ ,
\end{align}
where $T^a_{n+k/m} = T^a e^{i \frac{x^6}{R} (n+\frac{k}{m})}$. The commutation relation of new basis is given by 
\begin{align}
\comm{T^a_{n+k/m}}{T^b_{n'+k'/m}} = f^{abc} T^c_{n+n' + (k+k')/m}\ .
\end{align}
If $\sigma$ is a conjugation, it is called an {\it inner automorphism}. In this case, the resultant algebra is the same as untwisted affine Lie algebra \cite{Fuchs:1992nq}. If $\sigma$ is not an inner automorphism, it is called an {\it outer automorphism} and the resultant algebra becomes different from untwisted affine Lie algebras. The new algebra generated by $\{T^a_{n+m/l}\}$ with possible central extensions is called a  twisted affine Lie algebra. 

\begin{figure}
\centering
\begin{subfigure}{0.45\textwidth}
\centering
\includegraphics[scale=1]{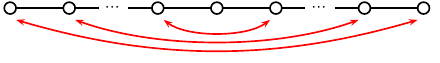}
\end{subfigure}
\hfill
\begin{subfigure}{0.45\textwidth}
\centering
\includegraphics[scale=1]{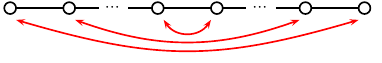}
\end{subfigure}
\begin{subfigure}{0.3\textwidth}
\centering
\includegraphics[scale=1]{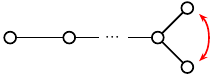}
\end{subfigure}
\hfill
\begin{subfigure}{0.3\textwidth}
\centering
\includegraphics[scale=1]{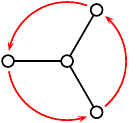}
\end{subfigure}
\hfill
\begin{subfigure}{0.3\textwidth}
\centering
\includegraphics[scale=1]{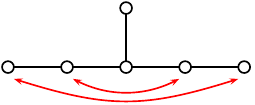}
\end{subfigure}
\caption{Dynkin diagrams and graph outer automorphisms of simple Lie algebras of type (a) $A_{2n+1}$, (b) $A_{2n}$, (c) $D_n$, (d) $D_4$ and (e) $E_6$.} \label{fig:dynkin_auto}
\end{figure}

\begin{figure}
\centering
\includegraphics[scale=1]{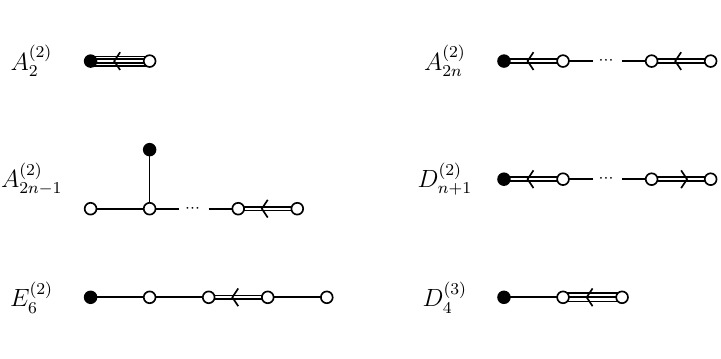}
\caption{Dynkin diagrams of twisted affine Lie algebras.} \label{fig:dynkin_twist}
\end{figure}

An outer automorphism can be viewed as a graph automorphism of Dynkin diagrams. Only the simple Lie algebras of types $A_n$, $D_n$ and $E_6$ have nontrivial outer automorphisms. $A_n$ has an $ \mathbb{Z}_2 $ Dynkin diagram automorphism which exchanges the simple roots $\alpha_i$ and $\alpha_{n+1-i}$ as in Figure~\ref{fig:dynkin_auto}(a) and Figure~\ref{fig:dynkin_auto}(b). $D_n$ has an order two outer automorphism which exchanges the simple roots $\alpha_{n-1}$ and $\alpha_n$ as in Figure~\ref{fig:dynkin_auto}(c). The exceptional algebra $E_6$ also has an order two outer automorphism shown in Figure~\ref{fig:dynkin_auto}(e). The $D_4$ algebra additionally has an order three automorphism as Figure~\ref{fig:dynkin_auto}(d). For these simple Lie algebras $X_\ell$, twisted affine Lie algebras associated with order $r = 2, 3$ diagram automorphism are denoted by $X_\ell^{(r)}$, and their Dynkin diagrams are given in Figure~\ref{fig:dynkin_twist}. Note that unlike the untwisted case, excluding the affine node does not yield $ X_\ell $ algebra. For example, deleting the affine node from $ A_{2\ell}^{(2)} $ and $ D_{\ell+1}^{(2)} $ algebras give the Dynkin diagram of $ C_\ell $ and $ B_\ell $ algebras, respectively.


Since a 6d gauge theory compactified on a circle naturally has an affine Lie algebra structure, its perturbative spectrum can be read off from the representation theory of Lie algebras. One can write down the perturbative part of the partition function by using either affine root system or by decomposing the representation of 6d gauge algebra into the representation of the invariant subalgebra under the automorphism. For more details, see \cite{Kac:1990gs,Kim:2004xx} and also  Appendix~A of \cite{Kim:2020hhh}.

We list some relevant results of the perturbative part of the partition function $Z_{\rm pert}$, used in this paper.\\
\noindent $(i)$   
\underline{The untwisted compactification:} for a 6d gauge theory with gauge algebra $ \mathfrak{g} $, the W-boson contribution to the perturbative part is given by, in the unrefined limit, 
\begin{align}
Z_{\mathrm{pert}} = \operatorname{PE}\qty[\frac{2g}{(1-g)^2} \frac{1}{1-q} \qty(\chi_{\Delta_+}^{\mathfrak{g}} + q \,\chi_{\Delta_-}^{\mathfrak{g}}) ]\ ,
\end{align}
where $\chi_{\Delta_\pm}^{\mathfrak{g}}$ are the character associated with the positive/negative roots, respectively. 
The twisted compactifications that we discussed in the main text are $SO(8)$ gauge theory with $\mathbb{Z}_2$ twist and $SU(3)$ gauge theory with $\mathbb{Z}_2$ twist. Their invariant subalgebras carry fractional KK-momentum. These fractional momenta contribute to the perturbative part. \\ 

\noindent $(ii)$ 
\underline{$ SO(8) $ gauge theory with $ \mathbb{Z}_2 $ twist:} the adjoint representation of $ \mathfrak{so}(8) $ decomposes into the adjoint and the fundamental representations of $ \mathfrak{so}(7) $. The adjoint representation of $ \mathfrak{so}(7) $ carries integer KK charge, while the fundamental representation carries half-integer KK charge,  
\begin{align}
D_4 &\to B_3 \nonumber \\
\mathbf{28} &\to \mathbf{21}_{0} \oplus \mathbf{7}_{1/2}\ , 
\end{align}
where the subscript denotes the KK charge. From this, one finds that the perturbative contribution to  the partition function for $ SO(8) $ gauge theory with $ \mathbb{Z}_2 $ twist is given by
\begin{align}
Z_{\mathrm{pert}}
= \operatorname{PE}\qty[\frac{2g}{(1-g)^2} \frac{1}{1-q} \qty( \chi_{\Delta_+}^{\mathfrak{so}(7)} + q^{1/2} \chi_{\mathbf{7}}^{\mathfrak{so}(7)} + q \,\chi_{\Delta_-}^{\mathfrak{so}(7)} ) ] \, .
\end{align}
In the dimensional reduction limit where $R \to 0 $, all the states with non-zero KK-momentum are truncated so that only the adjoint representation of $ \mathfrak{so}(7) $ algebra remains. We note that the invariant subalgebra $ \mathfrak{so}(7) $ can be readily read off, as it it nothing but the algebra obtained by removing the affine node of the Dynkin diagram of $ D_4^{(2)} $.\\

\noindent $(ii)$ \underline{$ SU(3) $ gauge theory with $\mathbb{Z}_2$ twist:} the adjoint representation of $ \mathfrak{su}(3) $ decomposes into the representations of $ \mathfrak{su}(2) $ as 
\begin{align}
A_2 &\to A_1 \nonumber \\
\mathbf{8} &\to \mathbf{3}_0 \oplus \mathbf{2}_{1/4} \oplus \mathbf{2}_{3/4} \oplus \mathbf{1}_{1/2} \, .
\end{align}
The perturbative contribution to the partition function for $ SU(3) $ gauge theory with $ \mathbb{Z}_2 $ twist is then given by
\begin{align}
Z_{\mathrm{pert}}
= \operatorname{PE}\qty[ \frac{2g}{(1-g)^2} \frac{1}{1-q} \qty(\chi_{\Delta_+}^{\mathfrak{su}(2)} + (q^{1/4}+q^{3/4}) \chi_{\mathbf{2}}^{\mathfrak{su}(2)} + q \, \chi_{\Delta_-}^{\mathfrak{su}(2)}) ]\ , 
\end{align}
up to the Cartan part. Again, in dimensional reduction limits, only the adjoint representation of $ \mathfrak{su}(2) $ survives, and this $ \mathfrak{su}(2) $ is the algebra obtained by removing the  affine node of the Dynkin diagram of $ A_2^{(2)} $.

\section{Special functions} \label{sec:special}

The Plethystic exponential is defined by
\begin{align}
\operatorname{PE}[f(x)] = \exp(\sum_{n=1}^\infty \frac{1}{n} f(x^n))\ .\nonumber
\end{align}
Its inverse function, Plethystic logarithm, is given by
\begin{align}
\operatorname{PL}\qty[f(x)] =\operatorname{PE}^{-1}\qty[f(x)] = \sum_{n=1}^\infty \frac{\mu(n)}{n} \log f(x^n)\ ,
\end{align}
where $\mu(n)$ is the M\"obius function defined by
\begin{align}
\mu(n) = \left\{
\begin{array}{ll}
(-1)^p &~ \text{if $n$ is a square-free positive integer with $p$ prime factors,} \\
0 &~\text{if~} n \text{ has a squared prime factor.}
\end{array}\right.
\end{align}

In topological vertex formalism, we use the following special functions for integer partitions $\lambda=(\lambda_{1},\lambda_{2},\cdots, \lambda_{\ell(\lambda)})$ and $\mu=(\mu_{1},\mu_{2},\cdots, \mu_{\ell(\mu)})$,
\begin{align}
\tilde{Z}_\nu(g) &= \tilde{Z}_{\nu^t}(g)
= \prod_{i=1}^{\ell(\nu)} \prod_{j=1}^{\nu_i} \qty(1 - g^{\nu_i + \nu_j^t - i - j + 1})^{-1}, \\
\mathcal{R}_{\lambda \mu}(Q) &= \mathcal{M}(Q)^{-1} \mathcal{N}_{\lambda^t \mu}(Q)\ , \\
\mathcal{M}(Q) &= \prod_{i, j=1}^{\infty} (1-Q g^{i+j-1})^{-1} = \operatorname{PE}\qty[\frac{g Q}{(1-g)^2}]\ , \\
\mathcal{N}_{\lambda\mu}(Q)
&= \qty[\prod_{i=1}^{\ell(\lambda)} \prod_{j=1}^{\lambda_i} \qty(1-Q g^{\lambda_i + \mu^t_j -i-j+1})] \qty[\prod_{i=1}^{\ell(\mu)} \prod_{j=1}^{\mu_i} \qty(1-Q g^{-\lambda^t_j-\mu_i +i+j-1})]\ ,
\end{align}
where $g$ the unrefined $\Omega$-deformation parameter, $\nu^t$ means transposed partition of  $\nu$, and $Q$ is a K\"ahler parameter. 
In practical calculation, the following Cauchy identities \cite{macdonald_symmetric_1995} are also handy:
\begin{align}
\sum_\lambda Q^{\abs{\lambda}} s_{\lambda/\eta_1}(\mathbf{x}) s_{\lambda/\eta_2}(\mathbf{y})
&= \prod_{i, j} \frac{1}{1 - Q x_i y_j} \sum_\lambda Q^{\abs{\lambda}} s_{\eta_2/\lambda}(\mathbf{x}) s_{\eta_1/\lambda}(\mathbf{y}) \label{eq:cauchy} \ , \\
\sum_\lambda Q^{\abs{\lambda}} s_{\lambda/\eta_1^t}(\mathbf{x}) s_{\lambda^t/\eta_2}(\mathbf{y}) &= \prod_{i, j} \qty(1+ Qx_i y_j) \sum_{\lambda} Q^{\abs{\lambda}} s_{\eta_2^t/\lambda}(\mathbf{x}) s_{\eta_1/\lambda^t}(\mathbf{y}) \label{eq:cauchy'}\ .
\end{align}
When $\mathbf{x} = g^{-\rho - \nu_1}$ and $\mathbf{y}=g^{-\rho - \nu_2}$ for $\rho = (-\frac{1}{2}, -\frac{3}{2}, -\frac{5}{2}, \cdots)$, as discussed in the main text,
\begin{align}
&\sum_\lambda Q^{\abs{\lambda}} s_{\lambda/\eta_1}(g^{-\rho-\nu_1}) s_{\lambda/\eta_2}(g^{-\rho-\nu_2}) 
= \mathcal{R}_{\nu_2 \nu_1}(Q)^{-1} \sum_\lambda Q^{\abs{\eta_1}+\abs{\eta_2}-\abs{\lambda}} s_{\eta_2/\lambda}(g^{-\rho-\nu_1}) s_{\eta_1/\lambda}(g^{-\rho-\nu_2}), \nonumber\\
&\sum_\lambda Q^{\abs{\lambda}} s_{\lambda/\eta_1^t}(g^{-\rho-\nu_1}) s_{\lambda^t/\eta_2}(g^{-\rho-\nu_2}) 
= \mathcal{R}_{\nu_2 \nu_1}(-Q) \sum_\lambda Q^{\abs{\eta_1}+\abs{\eta_2}-\abs{\lambda}} s_{\eta_2^t/\lambda}(g^{-\rho-\nu_1}) s_{\eta_1/\lambda^t}(g^{-\rho-\nu_2}).\nonumber
\end{align}
We also note that skew Schur function $s_{\lambda/\mu}$ is zero unless $\lambda \supset \mu$.\\


For a periodic strip diagram given in Figure~\ref{fig:6d_SU3Z2_LR}, 
the explicit form of such periodic vertex can be obtained using the method in \cite{Haghighat:2013gba, macdonald_symmetric_1995}. Using the definition of the edge factor and the vertex factor, one needs to evaluate the equation of the form
\begin{align}
\mathcal{G}(\mathbf{x}_1,\! \cdots\!, \mathbf{x}_4, \mathbf{y}_1,\! \cdots\!, \mathbf{y}_4) 
&= \sum_{\lambda_i, \eta_i} q_1^{\abs{\lambda_1}} q_2^{\abs{\lambda_2}} q_3^{\abs{\lambda_3}} q_4^{\abs{\lambda_4}} s_{\lambda_1/\eta_1}(\mathbf{x}_1) s_{\lambda_1/\eta_4}(\mathbf{y}_1) s_{\lambda_2/\eta_1}(\mathbf{x}_2) \nonumber \\
&~~ \times  s_{\lambda_2/\eta_2}(\mathbf{y}_2) s_{\lambda_3/\eta_2}(\mathbf{x}_3) s_{\lambda_3/\eta_3}(\mathbf{y}_3) s_{\lambda_4/\eta_3}(\mathbf{x}_4) s_{\lambda_4/\eta_4}(\mathbf{y}_4).
\end{align}
A successive use of the Cauchy identities \eqref{eq:cauchy} yields 
\begin{align}
\mathcal{G}(\mathbf{x}_1, \cdots, \mathbf{y}_4)
= \mathcal{F}(\mathbf{x}_1, \cdots, \mathbf{y}_4) \mathcal{G}(Q \mathbf{x}_1, \cdots, Q \mathbf{x}_4, Q \mathbf{y}_1, \cdots, Q \mathbf{y}_4) ,
\end{align}
where $Q = q_1 q_2 q_3 q_4$ and $\mathcal{F}(\mathbf{x}_1, \cdots, \mathbf{y}_4) = F(\mathbf{x}_1, \cdots, \mathbf{y}_4) F(Q^{1/2}\mathbf{x}_1, \cdots, Q^{1/2}\mathbf{y}_4)$ for
\begin{align}
&F(\mathbf{x}_1, \cdots, \mathbf{y}_4)   \\
& = \prod \Bigg[ \frac{1}{\left(1-q_1 \mathbf{x}_1 \mathbf{y}_1\right) \left(1-q_2 \mathbf{x}_2 \mathbf{y}_2\right) \left(1-q_3 \mathbf{x}_3 \mathbf{y}_3\right) \left(1-q_4 \mathbf{x}_4 \mathbf{y}_4\right) \left(1-q_{1,2} \mathbf{y}_1 \mathbf{y}_2\right) } \nonumber \\
&\times \frac{1}{\left(1-q_{1,4} \mathbf{x}_1 \mathbf{x}_4\right) \left(1-q_{2,3} \mathbf{x}_2 \mathbf{y}_3\right) \left(1-q_{3,4} \mathbf{x}_3 \mathbf{y}_4\right) \left(1-q_{1,2,3} \mathbf{y}_1 \mathbf{y}_3\right) \left(1-q_{1,2,4} \mathbf{y}_2 \mathbf{x}_4 \right) } \nonumber \\
&\times \frac{1}{\left(1-q_{1,3,4} \mathbf{x}_3 \mathbf{x}_1\right) \left(1-q_{2,3,4} \mathbf{x}_2 \mathbf{y}_4\right) \left(1- Q \mathbf{x}_2 \mathbf{x}_1\right) \left(1-Q \mathbf{y}_2 \mathbf{x}_3\right) \left(1-Q \mathbf{x}_4 \mathbf{y}_3\right) \left(1-Q \mathbf{y}_1 \mathbf{y}_4\right)} \Bigg].\nonumber
\end{align} 
Here, $q_{i,j,\cdots} = q_i q_j \cdots$ and $\prod (1-Q \mathbf{x}_i \mathbf{y}_j) = \prod_{m, n} \big(1-(\mathbf{x}_i)_m (\mathbf{y}_j)_n Q\big)$ for integer partitions $\mathbf{x}_i$ and $\mathbf{y}_j$. After repeating this procedure for $m$ 
times, we need to change $\mathbf{x}_i$ and $\mathbf{y}_j$ to $Q^m \mathbf{x}_i$ and $Q^m \mathbf{y}_j$, respectively, and hence, 
\begin{align}
\mathcal{G}(\mathbf{x}_1, \cdots, \mathbf{y}_4) = \qty[\prod_{m=0}^n \mathcal{F}(Q^m \mathbf{x}_1, \cdots, Q^m \mathbf{y}_4)] \mathcal{G}(Q^{n+1} \mathbf{x}_1, \cdots, Q^{n+1} \mathbf{y}_4).
\end{align}
Under the condition $Q \to 0$ as $n \to \infty$ which is used for deriving the Cauchy identities, the only contributing terms in $\mathcal{G}(Q^{m} \mathbf{x}_1, \cdots, Q^{m} \mathbf{y}_4)$ 
are $\lambda_i = \eta_j$ for all $i$ and $j$:
\begin{align}
\lim_{n\to \infty} \mathcal{G}(Q^n \mathbf{x}_1, \cdots, Q^n \mathbf{y}_4)
= \sum_\lambda Q^{\abs{\lambda}} = \prod_{m=1}^{\infty} \frac{1}{1-Q^m}\ . 
\end{align}
Hence, it follows that 
\begin{align}\label{eq:periodic_identity}
\mathcal{G}(\mathbf{x}_1, \cdots, \mathbf{y}_4)
= \prod_{n=1}^{\infty} \frac{\mathcal{F}(Q^{n-1} \mathbf{x}_1, \cdots, Q^{n-1} \mathbf{y}_4)}{1-Q^{n}}\ .
\end{align}

\vspace{0.5cm}
In localization computation, the Dedekind eta function $\eta$ and the Jacobi theta function $\theta_1(x)$ are defined as follows: For the complex structure $\tau$ of a torus,  $q = e^{2\pi i \tau}$,  
\begin{align}
\eta &= q^{\frac1{24}} \prod_{n=1}^\infty (1-q^n)\ , \\
\theta_1(x) &= -i q^{\frac18} y^{\frac12} \prod_{n=1}^{\infty} (1-q^n) (1- y q^n) (1-y^{-1} q^{n-1})\ ,
\end{align}
where $y=e^{2\pi i x}$. 
They satisfy
\begin{align}
&\theta_1(-x) = -\theta_1(x), \quad \theta_1(n \tau) = 0 \quad (n \in \mathbb{Z})\ , \nonumber \\
&\frac{1}{2\pi i} \oint_{u=0} \frac{du}{\theta_1(u)} = \frac{1}{2\pi \eta^3}\ , 
\end{align}
where $\oint_{u=0}$ means that the integral contour is taken around $u=0$ so that only the residue at $u=0$ contributes. 
These properties 
are useful when we evaluate the JK-residue. \\
%

\bibliographystyle{JHEP}
\bibliography{ref}
\end{document}